\documentclass[11pt]{article}
\usepackage[left=2.3cm, right=2.3cm, top=3cm]{geometry}
\usepackage{bm}
\usepackage{color}
\def\bSig\mathbf{\Sigma}

\usepackage{amsmath}
\usepackage{amssymb}
\usepackage{amsthm}
\usepackage[ruled,vlined]{algorithm2e}

\usepackage{float}

\usepackage[hidelinks,colorlinks=true,linkcolor=blue,citecolor=blue]{hyperref}

\usepackage{graphicx}

\usepackage{natbib}
\usepackage{booktabs}
\usepackage{amsfonts}
\usepackage{amssymb}
\usepackage{dsfont}
\usepackage{eucal}
\usepackage{multirow}
\usepackage[dvipsnames]{xcolor}
\usepackage{caption}
\def\bSig\mathbf{\Sigma}

\title{\LARGE Frequency Selection in Bayesian Spectral Modeling of Time Series Data with Applications to Wearable Device Measurements}

\author{
  Beniamino Hadj-Amar\thanks{University of South Carolina, Department of Epidemiology and Biostatistics, Columbia, SC \& Department of Statistics, Rice University, Houston, TX. \texttt{hadjamar@mailbox.sc.edu}} \and
  Vaishnav Krishnan\thanks{Department of Neurology, Neuroscience, and Psychiatry \& Behavioral Sciences, Baylor College of Medicine, Houston, TX. \texttt{vaishnav.krishnan@bcm.edu}} \and
  Marina Vannucci\thanks{Department of Statistics, Rice University, Houston, TX. \texttt{marina@rice.edu}}
}

\begin{document}

\maketitle
\def\spacingset#1{\renewcommand{\baselinestretch}%
{#1}\small\normalsize} \spacingset{1}

\spacingset{1.42} 









		\begin{abstract} This paper introduces a Bayesian spike-and-slab framework for spectral analysis of time series data. The proposed method combines frequency selection and dimensionality reduction with a refined grid of candidate frequencies, enabling high-resolution recovery of oscillatory components while promoting sparsity through a structured spike-and-slab prior. A stochastic search algorithm efficiently explores the posterior space, yielding posterior inclusion probabilities that quantify the relevance of each frequency. We extend the framework to multivariate signals via a hierarchical prior on frequency inclusion patterns, allowing the model to capture both shared and component-specific rhythms across multiple time series. Extensive simulation studies demonstrate the method’s robustness and superior performance in frequency estimation and spectral power reconstruction compared to existing approaches. Applied to actigraphy data from individuals with partial-onset seizures, the univariate model identifies clinically relevant circadian and ultradian rhythms. In a second application, for the joint analysis of physical activity and skin temperature from a healthy individual, the multivariate model reveals partially overlapping rhythmic components consistent with known physiological coupling. This work establishes a powerful and interpretable approach to spectral analysis, with broad applicability to wearable data, chronobiology, and personalized health monitoring.
		\end{abstract}

%

\vspace{0.5em}



%

\section{Introduction}
\label{sec:intro}

Understanding the periodic patterns in cyclical phenomena provides critical insights into the rhythmic, and potentially predictable drivers of variability. Assessments of periodicity in a variety of measurable biological signals can enhance patient care by forecasting physiological events \citep{grant2020ultradian} and/or adverse events \citep{friedrichs2024seizure, vieluf2024ultradian}. Among a plethora of such biological signals that can be derived from wearable devices, continuously collected accelerometry at the wrist (``wrist actigraphy'') has received the greatest emphasis. Actigraphy offers a unique, non-invasive view into an individual's rest-activity cycles by capturing metrics related to sleep duration, timing, and daytime movement levels. In medical research, this data can reveal patterns that align with changes in seizure frequency and the prevalence of psychiatric symptoms, such as depression or anxiety \citep{lau2024sleep} and also correlate with aspects of neurological disability, such as seizure frequency and side effect burden \citep{abboud2023actigraphic}.  Increasingly, actigraphy is being complemented by additional wearable-derived signals—such as skin temperature, heart rate, or electrodermal activity—that provide orthogonal yet physiologically related information. When analyzed jointly, these modalities can reveal partially shared rhythms and enhance the detection of underlying chronobiological patterns, motivating the need for multivariate modeling frameworks.

Despite the promise of actigraphy, a challenge remains in accurately and automatically identifying the specific frequencies, amplitudes, and phases within activity data. These frequency components, acting as signatures within the data, may offer a novel biomarker for monitoring physiological conditions, potentially linking variations in rest-activity rhythms to neurological states. Current research thus demands a methodology capable of extracting these distinct periodicities directly from actigraphy time series, enabling more precise, data-driven interpretations of rest-activity dynamics. Moreover, multivariate signals—such as physical activity and skin temperature—can exhibit overlapping but distinct rhythms, motivating the need for joint modeling frameworks that capture shared and modality-specific dynamics. Robust, data-adaptive techniques for frequency selection are crucial to advance our ability to use actigraphy as a tool for continuous, real-world physiological monitoring in clinical settings.


Spectral analysis has long been a cornerstone for understanding stationary time series, with the periodogram being one of the earliest tools. Developed as the squared magnitude of the Fourier transform, the periodogram provides a raw estimate of the power spectral density but often requires smoothing as the estimate is not consistent \citep{brockwell1991time, shumway2000time}. Smoothing techniques, such as the Welch method and multi-taper spectral estimation, improve stability by averaging across overlapping segments or using orthogonal tapers \citep{percival1993spectral}. Fourier decomposition remains central, where time series are expressed as a sum of sinusoidal components. This framework underpins classical techniques like the discrete Fourier transform (DFT) and its variants, which form the basis for harmonic analysis. Representing a time series as the sum of sine and cosine functions at different frequencies provides a detailed and flexible framework for signal representation . However, including all Fourier frequencies can lead to several drawbacks \citep{tseng2020fourier, lange2021fourier}: (i) overfitting, where the model captures noise rather than meaningful periodic signals; (ii) computational scalability issues, as the number of coefficients grows with the length of the time series; and (iii) poor interpretability, as the model does not automatically identify the relevant frequencies that drive overall variation in the data. Hence, selecting the appropriate subset of frequencies is essential to balance model accuracy and computational efficiency.

Bayesian methods have been used to address challenges in spectral analysis by incorporating prior information and enabling probabilistic approaches to spectral estimation. For example, \citet{bretthorst1988excerpts} introduced a Bayesian framework for spectral analysis, applying it to univariate stationary processes through prior distributions on spectral densities. \citet{choudhuri2004bayesian} utilized Dirichlet process priors and Bernstein polynomials for flexible spectral density estimation. In the multivariate setting, Bayesian spectral modeling has been extended to capture shared and component-specific structure across multiple time series. For instance, \citet{cadonna2019bayesian} proposed a hierarchical framework that models the log-spectral density functions of multiple signals, allowing for borrowing of strength across series while retaining individual spectral characteristics. More recently, \citet{hu2023fast} developed computationally efficient methods for multivariate spectral estimation using smoothing priors, enabling the identification of common spectral features across high-dimensional signals with uncertainty quantification. However, these methods do not directly address the challenge of automatically selecting the significant peaks in the spectrum, and hence the frequencies that drive the variation in the data. More recently, \citet{granados2022brain} proposed a Bayesian nonparametric mixture of autoregressive kernels that can data-adaptively identify both spectral peak locations and their associated bandwidths. In the frequentist literature, penalized techniques like LASSO and Elastic Net \citep{tibshirani1996regression, zou2005regularization} may be applied in the context of a saturated Fourier regression, where all possible Fourier frequencies are initially included, to encourage sparsity and improve interpretability \citep{incremona2019spectral, ha2019performance}. However, while these methods can help simplify the model, they can also lead to overfitting by selecting more frequencies than necessary, making it harder to identify the true underlying periodic signals. 


In this article, we propose a Bayesian spectral modeling approach that uses discrete spike-and-slab priors for simultaneous frequency selection and dimensionality reduction. The model uses a bivariate spike-and-slab prior over sinusoidal coefficients, enabling joint frequency selection across sine and cosine terms. A minimum-separation constraint excludes near-redundant components, while frequency-specific inclusion probabilities allow adaptation to heterogeneous spectral features.  Our prior model is complemented by incorporating a refined grid of candidate frequencies, to improve the resolution of the model. Traditional methods relying on Fourier frequencies often lack the precision needed to capture true periodicities, especially when fine temporal details are critical. By introducing a finer grid of candidate frequencies, the model achieves more accurate recovery of periodic signals and enhanced spectral estimates. While this refinement increases the number of frequency candidates, it provides significant gains in precision and accuracy, as demonstrated in our simulation studies.  Together, the spike-and-slab prior and the refined frequency grid enable a powerful and flexible framework for spectral analysis. To model multivariate signals, we extend the univariate prior by introducing a hierarchical structure on frequency inclusion indicators that enables flexible learning of shared and component-specific periodicities across time series. By placing a Dirichlet prior on inclusion patterns, the model borrows strength across signals while allowing asynchronous rhythms. For posterior inference,
we implement a stochastic search algorithm that efficiently explores the space of frequency models by iteratively adding, deleting, or swapping frequencies. 
We further improve computational efficiency by focusing on the most relevant regions of the posterior space, guided by informed initialization based on dominant frequencies in the data. Our proposed spike-and-slab prior modeling framework allows to quantify posterior probabilities of inclusion for each frequency and the posterior distribution over the number of frequencies, offering insight into the model's complexity.


We evaluate the performance of our proposed approach through extensive simulation studies, demonstrating accuracy in identifying frequencies that drive variation in the data, robustness across diverse scenarios, and clear advantages over existing alternative methods. This includes both univariate and multivariate settings, with simulations illustrating the method’s ability to recover shared and component-specific rhythms across multiple time series. We then present two real-data applications. In the first, we apply the model to univariate actigraphy data collected from individuals with partial-onset seizures. We show that the model effectively identifies distinct frequency components by leveraging sparsity while preserving key oscillatory features. Specifically, the approach enables the robust extraction of circadian and ultradian rhythms, accurately capturing both dominant and subtle periodicities in the data. In the second application, we analyze bivariate data from a healthy individual, jointly modeling physical activity and skin temperature. This multivariate analysis reveals both shared and modality-specific components, capturing well-known physiological coupling between behavioral and thermoregulatory rhythms. By leveraging a structured spike-and-slab prior, the model promotes sparsity while flexibly capturing multiscale spectral structure.

The remainder of the paper is organized as follows: Section \ref{sec:Methods} introduces the model and prior specification, Section \ref{sec:MCMC_sampler} details the proposed MCMC algorithm and discusses posterior inference. Section \ref{sec:simulation_studies} presents the simulation studies, and Section \ref{sec:Data} illustrates the applications to measurements from wearable devices. Section \ref{sec:discussion} concludes the paper with a discussion.

\section{Methods}

\label{sec:Methods}

\subsection{Spectral Modeling}
\label{sec:model}

Consider a  time series realization $\bm{y} = (y_1, \dots, y_T)$, with $T$ denoting the length of the time series. For simplicity, we assume 
$T$ is even. In spectral analysis, the observed time series can be exactly expressed as a weighted sum of sines and cosines oscillating at the \textit{Fourier frequencies} $\tilde{\omega}_j$ = $j/T$, for $j=1, \dots, M_{max}$, with $ M_{max} = (T/2 -1)$ being the maximum number of frequencies included in the model. Hence,  we can formulate a saturated regression model involving all Fourier coefficients as
\begin{equation}
	y_t = \sum_{j=1}^{M_{\text{max}}} \Big\{ \beta_{j1} \cos(2\pi\tilde{\omega}_j t) + \beta_{j2} \sin{(2
		\pi \tilde{\omega}_j t)} \Big\} + \varepsilon_t,
	\label{eq:saturated_model}
\end{equation}
for $t=1, \dots, T$. Here, the linear coefficients $\beta_{j1}$ and $\beta_{j2}$ determine amplitude and phase of sine and cosine at frequency $\omega_j$; the power can be summarized as  $P(\omega_j) = \beta_{j1}^2 + \beta_{j2}^2$, which denotes the relative contribution of frequency $\omega_j$ to the overall variation in the data. Further, we assume normally distributed errors with zero mean and variance $\sigma^2$, i.e. $\varepsilon_t \sim \mathcal{N}(0, \sigma^2).$

\subsubsection{Dense Grid of Candidate Frequencies} Frequency domain analysis is inherently constrained by the duration of the input time series. According to the Fourier uncertainty principle, the resolution of frequency domain output improves with longer time series segments, as the number of available frequency bins increases with the temporal integration window \citep{percival1993spectral}. Typically, one would consider the set of Fourier frequencies, $\tilde{\omega}_j = j/T$, as potential candidates for inclusion in a model. However, this approach limits the precision with which one can recover true periodicities. For instance, in cases where fine temporal details are needed, relying solely on Fourier frequencies may result in estimates that are close but not precise.
To address this problem, we consider a dense grid of candidate frequencies,  $\bar{\omega}_j = j \cdot \Delta \omega$ for $j = 0, 1, 2, \dots, L$, where $\Delta \omega$ controls the grid resolution, and $L$ is chosen such that $\bar{\omega}_j \in (0, 0.5)$, since the contribution of each frequency is symmetric and periodic, so frequencies above $0.5$ are aliased with those below and do not provide additional information. By decreasing $\Delta \omega$, we enhance the frequency resolution, enabling more precise recovery of true periodicities. In section  \ref{sec:simulation_studies} we show on simulated data how a denser grid not only improves the estimate of the location of the frequencies but also their estimated power.

This choice provides a denser discretization of the frequency domain, which can improve the ability to localize true periodicities compared to approaches restricted to the Fourier frequencies. As shown in \citet{keil2022recommendations}, finer grids improve resolution, particularly in cases where high temporal resolution competes with the need for fine frequency distinctions. We note that \citet{wu2024frequency} also advocate the use of dense frequency grids with sufficiently fine mesh size to achieve improved frequency localization and near-parametric estimation accuracy in complex or non-stationary time series. By improving the grid resolution, we mitigate aliasing effects, as highlighted by the Nyquist sampling theorem, and achieve a more precise balance between temporal and frequency resolution \citep{keil2022recommendations}. However, it is important to note that as the grid becomes finer, the number of potential frequencies increases, leading to higher computational costs. Therefore, the more precise localization of oscillatory components comes with a trade-off in terms of computational efficiency.


\subsection{Frequency Selection through Spike-and-Slab Priors}
We propose an approach based on discrete spike-and-slab priors that allows for simultaneous frequency selection and dimensionality reduction. Spike-and-slab priors are commonly used in regression models to achieve variable selection with the goal of identifing the most relevant predictors from a set of available covariates \citep{mitchell1988bayesian, george1993variable, brown:1998}. These priors combine two components: a \textit{spike} that concentrates probability mass at zero (indicating that a variable is not included in the model) and a \textit{slab} that has its mass distributed over a broad range of possible values. In the regression context, these sparsity-inducing mixture priors are applied to the regression coefficients, allowing irrelevant covariates to be excluded by assigning their coefficients a value of zero. Spike-and-slab priors offer several benefits, including the ability to facilitate complex modeling through MCMC-based stochastic search, provide accurate predictions via model averaging, adapt seamlessly to multivariate and nonlinear frameworks, and address high-dimensional problems where the number of covariates surpasses the available data. Additionally, they enable the integration of prior knowledge into the modeling process. See \cite{tadesse2021handbook} for a comprehensive review of theoretical, methodological and computational aspects of these priors.  In our modeling framework, the spike part of the prior corresponds to assuming that some frequencies may not be relevant in the analysis; that is, some coefficients in the spectral decomposition (Eq. \ref{eq:saturated_model}) should be set to zeros. On the other hand, the slab part allows the inclusion of frequencies with non-zero coefficients, recognizing that certain frequencies may have a meaningful impact on the signal and should be included in the model.

Let $z_j$ be the latent inclusion indicator variable for frequency $\omega_j$, such that $z_j = 1$ if $\omega_j$ significantly contributes to driving the overall oscillatory behavior of the time series, and $z_j = 0$ otherwise. We assume a discrete spike-and-slab prior on the coefficients $\bm{\beta}_j = (\beta_{j1}, \beta_{j2})$ as
\begin{equation}
	\left[ \bm{\beta}_{j} \mid z_{j} \right] \sim z_{j} \cdot \mathcal{N}_2(\bm{0}, \sigma^2_\beta \bm{I}_2 ) + (1 - z_{j}) \cdot \bm{\delta}_0 (\bm{\beta}_{j}),
	\label{eq:spike_slab}
\end{equation}
where $\bm{\delta}_0$ denotes the bivariate Dirac mass at zero, that is, $\bm{\delta}_0(\bm{\beta}_j) = 1$ if $\bm{\beta}_j = \bm{0}$ and $0$ otherwise. 
The slab variance $\sigma^2_\beta$ is a hyperparameter controlling the prior dispersion of the active coefficients. In the applications of this paper, we fix $\sigma_\beta^2$ to a moderately large value, to avoid over-shrinkage. We enforce sparsity and improve interpretability by assuming a constrained joint prior on the inclusion vector $\mathbf{z} = (z_1, \dots, z_{M_{\max}})$ defined as
\begin{equation}
	p(\mathbf{z} \mid \bm{\alpha}) = \left[ \prod_{j=1}^{M_{\max}} \alpha_j^{z_j} (1 - \alpha_j)^{1 - z_j} \right] \cdot \mathbb{I}\left( \min_{j < k:\, z_j = z_k = 1} |\omega_j - \omega_k| \geq d \right),
	\label{eq:prior_z}
\end{equation}
where $\mathbb{I}(\cdot)$ is the indicator function and $d$ is a user-defined minimum frequency separation threshold. This constraint ensures that any pair of active frequencies in the model are separated by at least $d$, thereby reducing collinearity among basis elements and improving identifiability. Configurations violating this constraint receive zero prior probability and are excluded from the posterior support. A similar principle is discussed by \citet{hadj2020bayesian}. This minimum-separation rule reflects both modeling and computational considerations: nearby frequencies are often difficult to distinguish given finite sample sizes, and including them may lead to overfitting or unstable inference. By integrating the separation condition directly into the prior, we guarantee that all valid configurations reside on a well-behaved, interpretable, and identifiable subspace of the model space. This formulation leads to a parsimonious and structured representation that facilitates automatic frequency selection while preserving model coherence. See Section \ref{sec:selectin_minimum_d} below for further discussion on how to set the parameter $d$. Furthermore, we assume $\alpha_j \sim \text{Beta}(a, b)$ and integrate the $\alpha_j$'s out analytically, leading to a collapsed marginal prior on $z_j$ given by $p(z_j = 1) = \frac{a}{a + b}$ and  $p(z_j = 0) = \frac{b}{a + b}$, noting that the minimum-separation constraint remains in effect under the marginal prior, as configurations violating the threshold continue to receive zero probability.
By allowing frequency-specific inclusion probabilities $\alpha_j$ we achieve greater flexibility than using a single global parameter $\alpha$, as this allows the model to accommodate heterogeneity in the likelihood of inclusion across different regions of the frequency spectrum. We complete the prior model with an inverse-gamma prior on the residual variance, $\sigma^2 \sim \text{IG}(\gamma_0/2, \nu_0/2)$.

\subsubsection{Minimum Separation Constraint: Specification and Rationale} \label{sec:selectin_minimum_d}

As introduced in the prior in Eq.~\eqref{eq:prior_z}, the inclusion vector $\bm{z} = (z_1, \dots, z_{M_{\max}})$ is restricted to configurations in which any pair of included frequencies are separated by at least $d$ bins on the refined frequency grid. We now discuss the rationale for setting this parameter and its role in the model. In our implementation, we fix the minimum separation to $d = 5$, which corresponds to a frequency gap of approximately five bins on the refined grid. This choice reflects both biological plausibility and statistical stability. In the context of our applications to wearable data, we aim to capture dominant oscillations such as the circadian rhythm (approximately 24 hours), ultradian rhythms (e.g., 12 hours), and infradian cycles. On a fine frequency grid—such as one constructed with resolution $1/N$—adjacent bins may correspond to very similar periods. For instance, frequencies near the circadian range might correspond to periods of 24.64, 24.30, 23.96, 23.63, and 23.31 hours. Including both 23.96 and 23.31 hours in the same model would be physiologically implausible, as they reflect nearly indistinguishable periodicities from a biological standpoint. 
Formally, the separation constraint enforces the condition:
\[
\min_{j < k:\, z_j = z_k = 1} |\omega_j - \omega_k| \geq d \cdot \Delta\omega,
\]
where $\Delta\omega$ is the frequency bin width. This ensures that the selected frequency components $\{ \omega_j : z_j = 1 \}$ are well separated, reducing redundancy and improving identifiability of latent periodicities. In practice, this has two major benefits: (i) it avoids collinearity among the basis vectors associated with nearby frequencies, thereby improving the stability of posterior inference; and (ii) it concentrates posterior inclusion probabilities (PIPs) around dominant components. Without the constraint, the model may spread posterior mass across a cluster of collinear frequencies, making interpretation difficult and diluting the PIP of each.

Importantly, while $d$ is a user-specified parameter, its effects are interpretable and predictable. Smaller values of $d$ increase flexibility but may lead to overfitting or redundancy; larger values promote parsimony but may miss closely spaced true signals. In our application, we found $d = 5$ to provide a balanced compromise, yielding stable posterior summaries and interpretable rhythm estimates across a range of subjects. Note that as the frequency grid becomes more refined (i.e., with smaller $\Delta\omega$), a proportionally larger value of $d$ helps maintain identifiability by preventing the inclusion of highly correlated or physiologically indistinguishable components. The impact of $d$ in settings with closely spaced frequencies is further investigated in a simulation study reported in the Supplementary Material.

\subsection{Extension to Multivariate Data with Shared Spectral Structure} \label{sec:multivariate}
The univariate setting introduced earlier models an individual time series as a sum of sinusoidal components, with spike-and-slab priors guiding frequency selection. In multivariate settings, where multiple related time series are observed,  it is often reasonable to expect that some oscillatory components recur across datasets. Applying the univariate model to each time series separately would ignore these potential commonalities. We propose an extension to a joint model that 
retains component-specific flexibility for each time series while introducing a hierarchical prior to couple frequency inclusion across datasets.

Let $\mathbf{y}_t = (y_{t1}, \ldots, y_{tD})^\top \in \mathbb{R}^D$ denote the observed values of a $D$-dimensional time series at time $t = 1, \ldots, T$. For each component $i \in \{1, \ldots, D\}$, we model the time series as a linear combination of sinusoidal functions with component-specific coefficients:
\[
y_{ti} = \sum_{j=1}^{M_{\max}} \left\{ \beta_{ji}^{(1)} \cos(2\pi \omega_j t) + \beta_{ji}^{(2)} \sin(2\pi \omega_j t) \right\} + \varepsilon_{ti}, \quad \varepsilon_{ti} \sim \mathcal{N}(0, \sigma_i^2),
\]
where the set of candidate frequencies \( \omega_j \) has already been introduced in the univariate formulation, and \( \boldsymbol{\beta}_{ji} = (\beta_{ji}^{(1)}, \beta_{ji}^{(2)})^\top \) are the component-specific coefficients determining the contribution of each frequency to component \( i \). Here \( \sigma_i^2 \) is the residual variance for the \( i \)-th component. In order to extend the spike-and-slab prior to the multivariate setting, we introduce binary indicators $z_{ji} \in \{0,1\}$ for each frequency $\omega_j$ and component $i$, such that $\boldsymbol{\beta}_{ji}$ is drawn from a slab prior if $z_{ji}=1$ and set to zero otherwise. This allows each component of the multivariate signal to selectively include or exclude any frequency. Rather than treating each $z_{ji}$ independently, we define a joint inclusion pattern $\mathbf{z}_j = (z_{j1}, \ldots, z_{jD}) \in \{0,1\}^D$ across all components for each frequency $\omega_j$, as described below.  

\subsubsection{Structured Frequency Inclusion via Hierarchical Priors}

To introduce structured information sharing across components, we place a hierarchical prior on the joint inclusion pattern $\mathbf{z}_j$ for each frequency $\omega_j$. Specifically, we assume that $\mathbf{z}_j$ is drawn from a categorical distribution with probability vector $\boldsymbol{\pi}$, which itself is given a Dirichlet prior, e.g., 
$\mathbf{z}_j \sim \text{Categorical}(\boldsymbol{\pi})$ and $\boldsymbol{\pi} \sim \text{Dirichlet}(\boldsymbol{\alpha})$.
For example, in the bivariate case ($D=2$), the vector $\boldsymbol{\pi} = (\pi_{00}, \pi_{10}, \pi_{01}, \pi_{11})$ encodes the prior probabilities of all possible inclusion configurations across the two components. This hierarchical formulation couples frequency selection across components, allowing the model to borrow strength and favor synchronized inclusion patterns when supported by the data, while still permitting component-specific selections when necessary. For instance, a high value of $\pi_{11}$ increases the probability that the same frequency is selected in both components, reflecting potential physiological synchronization or common periodic regulation. On the other hand, higher values of $\pi_{10}$ or $\pi_{01}$ allow for frequency inclusion in only one component, thus preserving the capacity for component-specific rhythms. By inferring $\boldsymbol{\pi}$ from the data, the model can adaptively learn whether spectral patterns are shared or distinct across components.  Furthermore, the Dirichlet prior on $\boldsymbol{\pi}$ introduces further modeling flexibility by enabling us to encode prior beliefs about the prevalence of different patterns. Specifically, the hyperparameters $\boldsymbol{\alpha} = (\alpha_{00}, \alpha_{10}, \alpha_{01}, \alpha_{11})$ control the prior weight placed on each configuration: larger values of $\alpha_{00}$ promote sparsity by favoring exclusion from both components, while higher values of $\alpha_{11}$ encourage joint inclusion. This allows the prior to reflect different expectations about frequency sharing based on domain knowledge or modeling goals.

As in the univariate setting, we impose the same minimum-separation constraint on the selected frequencies within each component. Specifically, for any two frequencies $\omega_j$ and $\omega_k$ included in the same component $i$, we require that they be separated by at least $d$ units on the frequency grid. This constraint is incorporated into the joint prior as:
\[
p(\{\mathbf{z}_j\}_{j=1}^{M_{\max}} \mid \boldsymbol{\pi}) \propto \left[ \prod_{j=1}^{M_{\max}} \pi(\mathbf{z}_j) \right] \cdot \mathbb{I}\left( \min_{\substack{j < k; z_{ji} = z_{ki} = 1}} |\omega_j - \omega_k| \geq d \ \text{for all } i \in \{1, \ldots, D\} \right).
\]
This enforces the separation constraint in a component-wise manner, ensuring that active frequencies remain well-separated in each dimension without redundancy.

While the inclusion indicators $\mathbf{z}_j$ govern whether a frequency is selected in each component, the model retains flexibility at the coefficient level. Specifically, when $z_{ji} = 1$, the corresponding component $i$ maintains its own coefficient vector  $\boldsymbol{\beta}_{ji}$, allowing for differences in amplitude and phase across components even when the same frequency is shared. This formulation captures both synchrony and heterogeneity: shared frequencies can exhibit aligned periodic behavior across components, yet magnitude and phase of oscillations may vary. 
This allows the model to support flexible and interpretable multivariate spectral decompositions. 

\section{Markov Chain Monte Carlo Sampler}
\label{sec:MCMC_sampler}

We first describe the MCMC sampler for the univariate model of Section~\ref{sec:model}. Let $\bm{\beta} = (\bm{\beta}_1^{'}, \dots, \bm{\beta}_{M_{max}}^{'})^{'}$ be the vector of coefficients corresponding to the entire set of frequencies $\bm{\omega} = (\tilde{\omega}_1, \dots, \tilde{\omega}_{M_{max}})$, where $\bm{\beta}_j = (\beta_{j1}, \beta_{j2})^{'}$,  and let $\bm{z} = (z_1, \dots, z_{M_{max}})$ denote the vector of inclusion indicators for the frequencies. Then, Bayesian inference is based on the following factorization of the joint posterior distribution,

\begin{equation}
	p (\bm{z}, \bm{\beta}, \sigma^2 | \bm{y}, \bm{\omega}) \propto  p (\bm{y} | \bm{z}, \bm{\beta}, \sigma^2, \bm{\omega}) p(\bm{z}) p(\bm{\beta_z} | \bm{z}) p(\sigma^2). 
\end{equation}
We devise a Metropolis-within-Gibbs sampler that iteratively alternates between sampling the indicators $\bm{z}$, the variance $\sigma^2$ and the active coefficients $\bm{\beta}_{\bm{z}}$ corresponding to the included frequencies, namely $\bm{\beta}_{\bm{z}} = \{ \bm{\beta}_j : z_j = 1, \,\,
j =1, \dots, M_{max} \}$ for $\bm{\omega}_{\bm{z}} = \{ \tilde{\omega}_j : z_j = 1, \,\,
j =1, \dots, M_{max} \}$.  At a generic MCMC iteration we update the parameters as follows: 
\vspace{0.1cm}

\textbf{$\bullet$ Jointly updating active frequency indicators $\bm{z}$ and linear basis coefficients $\bm{\beta_z}$}: We perform a joint sampling of $\bm{z}$ and $\bm{\beta_z}$ following the frequently employed stochastic search MCMC algorithm of \cite{savitsky2011variable},  that cleverly avoids having to
	deal with the changing dimensions of the parameter space via a joint update of the inclusion indicators and the corresponding coefficients. 
This algorithm involves add-delete-swap moves, enabling the exploration of the posterior space by sequentially visiting models with different numbers of frequencies. 
Importantly, all proposed moves are embedded within a Metropolis--Hastings (MH) framework with appropriate acceptance probabilities, ensuring that the resulting Markov chain satisfies detailed balance and leaves the posterior distribution invariant. At each step, the new model differs from the previous one by including and/or excluding one frequency. The indices for modification are selected uniformly at random among valid candidates, conditional on the minimum separation constraint
provided in Eq. \eqref{eq:prior_z}. Thus, the move set is defined on the support of the prior. To elaborate,  at each iteration the new model is created from the previous one by randomly selecting one of the following  moves:

\begin{enumerate}
	\item \textit{Adding or deleting a frequency}: Select one of the $M_{max}$ indices in $\bm{z}^{curr}$ at random and modify its value (add: $0 \rightarrow 1$; delete: $1 \rightarrow 0$). This either includes a new frequency in the model or deletes a frequency currently included. This move is carried out jointly for the pair of coefficients $\bm{\beta}_j = (\beta_{j1}, \beta_{j2})$, that is,  they are simultaneously added or removed, hence ensuring that the corresponding frequency $\omega_j$ is added or removed, respectively, from the model. When adding a frequency, the new pair of coefficients $\bm{\beta}_j^{prop}$ is sampled from the conjugate Normal posterior distribution (see Eq. \ref{eq:conjugate_update} below) of the entire vector $\bm{\beta_z}$ of active linear coefficients, conditional on the proposed active frequencies determined by $\bm{z}^{prop}$. The $j$-th components are then selected as the proposed linear coefficients. The acceptance probability for adding a frequency is given by
	\begin{equation}
		\text{min} \bigg[1,  \dfrac{ p (\bm{y} | \bm{z}^{prop}, \bm{\beta}^{prop}, \sigma^2, \bm{\omega}) p(z_j^{prop}) \prod_{i=1}^2 p(\beta_{ji}^{prop} | z_j^{prop} ) }{p (\bm{y} | \bm{z}^{curr}, \bm{\beta}^{curr}, \sigma^2, \bm{\omega}) p (z_j^{curr})}\bigg],
		\label{eq:acceptance_prob}
	\end{equation}
	whereas the one for deleting a frequency is calculated similarly by inverting Eq. \eqref{eq:acceptance_prob}.
	
	\item \textit{Swapping frequencies}: Select a 0 and a 1 at random from $\bm{z}^{curr}$, then switch their values. This process adds a new frequency to the model while removing a currently included one. Naturally, this is reflected in the corresponding pair of coefficients $(\beta_{j1}, \beta_{j2})$.
\end{enumerate}

\textbf{$\bullet$ Updating the active linear basis coefficients \(\bm{\beta_z}\)}:  We perform a standard Bayesian update of the active linear coefficients \(\bm{\beta_z}\). This update is not required for ergodicity, but it helps to speed up the convergence of the chain by supplying relatively more sampled values of the parameters at a given iterations \citep{savitsky2011variable}. Specifically, the posterior distribution of \(\bm{\beta_z}\) conditional on \(\sigma^2\), \(\bm{y}\), and \(\bm{\omega_z}\) follows a multivariate normal distribution, \(\bm{\beta_z} | \sigma^2, \bm{y}, \bm{\omega_z} \sim \mathcal{N}_{2M}(\bm{\lambda}_T, \bm{\Sigma}_T)\), where $M$ is the number of active frequencies. The parameters of these distributions are defined as:
\begin{equation}
	\label{eq:conjugate_update}
	\bm{\Sigma}_T^{-1} =  \frac{1}{\sigma^2_\beta}\bm{I}_{2M} + \frac{1}{\sigma^2} \bm{X(\omega_z)}' \bm{X(\omega_z)}, \quad 
	\bm{\lambda}_T = \frac{1}{\sigma^2} \bm{\Sigma}_T^{-1} \bm{X(\omega_z)}' \bm{y}.
\end{equation}
Here, the spectral matrix $\bm{X}(\bm{\omega}_{\bm{z}})$ is the matrix of basis functions with rows given by 
$$\bm{x}_{t}(\bm{\omega}_{\bm{z}}) = \big(\cos(2\pi\tilde{\omega}_{1}), \sin(2\pi\tilde{\omega}_{1}), \dots, \cos(2\pi\tilde{\omega}_{M}), \sin(2\pi\tilde{\omega}_{M})\big), ~~t=1, \dots, T.$$

\textbf{$\bullet$ Updating the residual variance \(\sigma^2\)}: This is standard, as the posterior distribution of $\sigma^2$ follows an inverse-gamma distribution, \(\sigma^2 | \bm{y}, \bm{\beta_z}, \bm{\omega_z} \sim IG(\gamma_T, \nu_T)\), with updated parameters
\begin{equation}
	\gamma_T = \frac{T+\gamma_0}{2}, \quad \nu_T = \frac{1}{2}\left( \nu_0 + \sum_{t=1}^T  \{ y_t - \bm{x}_t\bm{(\omega_z)}' \bm{\beta_z} \}^2\right).
\end{equation}

As noted above, the minimum-separation constraint is not merely imposed on the proposal mechanism but is embedded in the prior (Eq. \ref{eq:prior_z}). Therefore, the posterior is zero outside this constrained support, and the sampler only explores valid configurations. Since the proposal mechanism ensures that each transition remains within this support and the move set connects all valid configurations (e.g., from $(1,0,0,\dots)$ to $(1,0,\dots,1)$ is reachable through a finite number of add/delete steps that maintain spacing), the Markov chain is irreducible over the support of the posterior. Hence, the sampler is valid and converges to the correct target distribution.  In our implementation, the list of valid candidate indices for add/delete and swap moves is dynamically updated at each MCMC step to reflect the current configuration and separation constraint. This ensures the proposal mechanism remains efficient and valid without revisiting zero-probability states.
Furthermore, to improve efficiency of the sampler and expedite convergence, 
we propose initializing the frequencies in the model using the locations of the top $M_{start}$ peaks from the periodogram. This approach leverages the periodogram as a simple yet effective tool for identifying dominant frequencies in the data, providing informed starting values and reducing time spent by the sampler exploring low-probability areas. 

\subsection{MCMC Sampler for Multivariate Data}
\label{sec:MCMC_multi}

The sampler scheme described above can be extended to the the case of multiple time series. In the setting described in Section \ref{sec:multivariate} each component retains its own set of Fourier coefficients and residual variance and inference is coupled across series through a shared prior over joint frequency inclusion patterns. Accordingly, the sampler enables (i) structured cross-component frequency selection, (ii) flexible reallocation across dimensions, and (iii) Bayesian learning of inclusion structure. We provide a short description of the algorithm here, with full details in the Supplementary Material.

At each iteration, the sampler proposes a local modification to the binary inclusion vector associated with a single frequency—either by toggling inclusion in one component (add/delete) or by swapping the inclusion status between two components, which may belong to the same or different dimensions. Crucially, these swap moves can reallocate spectral features across different components, allowing the sampler to explore cross-series spectral sharing in a flexible, data-driven way. Each proposal is accepted only if it respects a minimum-separation constraint within each component, and the MH ratio is computed using only the components affected by the change, which significantly improves computational efficiency. In line with the univariate sampler, each proposed inclusion move is paired with a joint update of the corresponding Fourier coefficients, sampled from their conditional Gaussian distribution under the proposed model. This pair \( (\mathbf{z}_j^{\text{prop}}, \boldsymbol{\beta}_j^{\text{prop}}) \) is then accepted or rejected using a MH step. After the inclusion indicators are updated, we perform a second stage of updates: we re-sample all active Fourier coefficients and component-specific residual variances from their full conjugate posteriors. This two-step strategy improves mixing by refining the coefficients under the current model configuration. Finally, the parameter of the Dirichlet prior over inclusion configurations is updated via its conjugate posterior, allowing the model to adaptively learn the relative prevalence of inclusion patterns across components. This hierarchical structure encourages shared frequency selection when supported by the data, while preserving the flexibility to model component-specific spectral features.

\subsection{Posterior Inference} \label{sec:posterior_inference}

The procedures described in this section apply to both the univariate and multivariate models introduced earlier. From the MCMC output, we estimate Posterior Probabilities of Inclusion (PPIs) for each frequency \( \omega_j \), formally defined as the posterior probability that \( z_j = 1 \), and  estimated from the MCMC samples as
$P(z_j = 1 \mid \bm{y}, \cdot) = \frac{1}{S} \sum_{s=1}^{S} z_j^{(s)}$,
with \( S \) the number of MCMC iterations and \( z_j^{(s)} \) the value of the latent indicator \( z_j \) at the \( s \)-th iteration. This quantity provides a probabilistic assessment of the importance of each frequency, allowing us to prioritize those with higher PPIs as key contributors to the signal. We follow  \citet{barbieri2004optimal}, and select key frequencies as those with PPI greater than 0.5. Alternatively, one could impose a more stricter rule based on high posterior inclusion probabilities (e.g., PPI $>0.9$) to promote additional parsimony in the model.
The model also provides a posterior distribution over the total number of selected frequencies, which serves as a natural measure of model complexity. This is computed as:
\[
P(M = j \mid \bm{y}, \cdot) = \frac{1}{S} \sum_{s=1}^{S} \mathbb{I}\left( \sum_{j=1}^{M_{\max}} z_j^{(s)} = j \right),
\]
where \( \mathbb{I}(\cdot) \) is the indicator function, which equals 1 if the sum of the latent indicators at iteration \( s \) is equal to \( j \), and 0 otherwise. In the multivariate setting, this distribution can be computed separately for each component, or jointly over inclusion patterns, depending on the structure of interest (see Section~\ref{sec:multivariate}). 

\section{Simulation Studies} \label{sec:simulation_studies}

We conduct simulation studies to assess the performance of our proposed method, hereafter referred to as \textbf{spectralSS} (spectral spike-and-slab).

\subsection{Illustrative Example} \label{sec:illustrative_example}
\begin{figure}[htbp]
	\centering
	\begin{tabular}{@{}c@{\hspace{0.5em}}c@{}}
		{\scriptsize (a)} &
		{\scriptsize (b)}
		\\[-1.2ex]
		\includegraphics[width=0.48\textwidth]{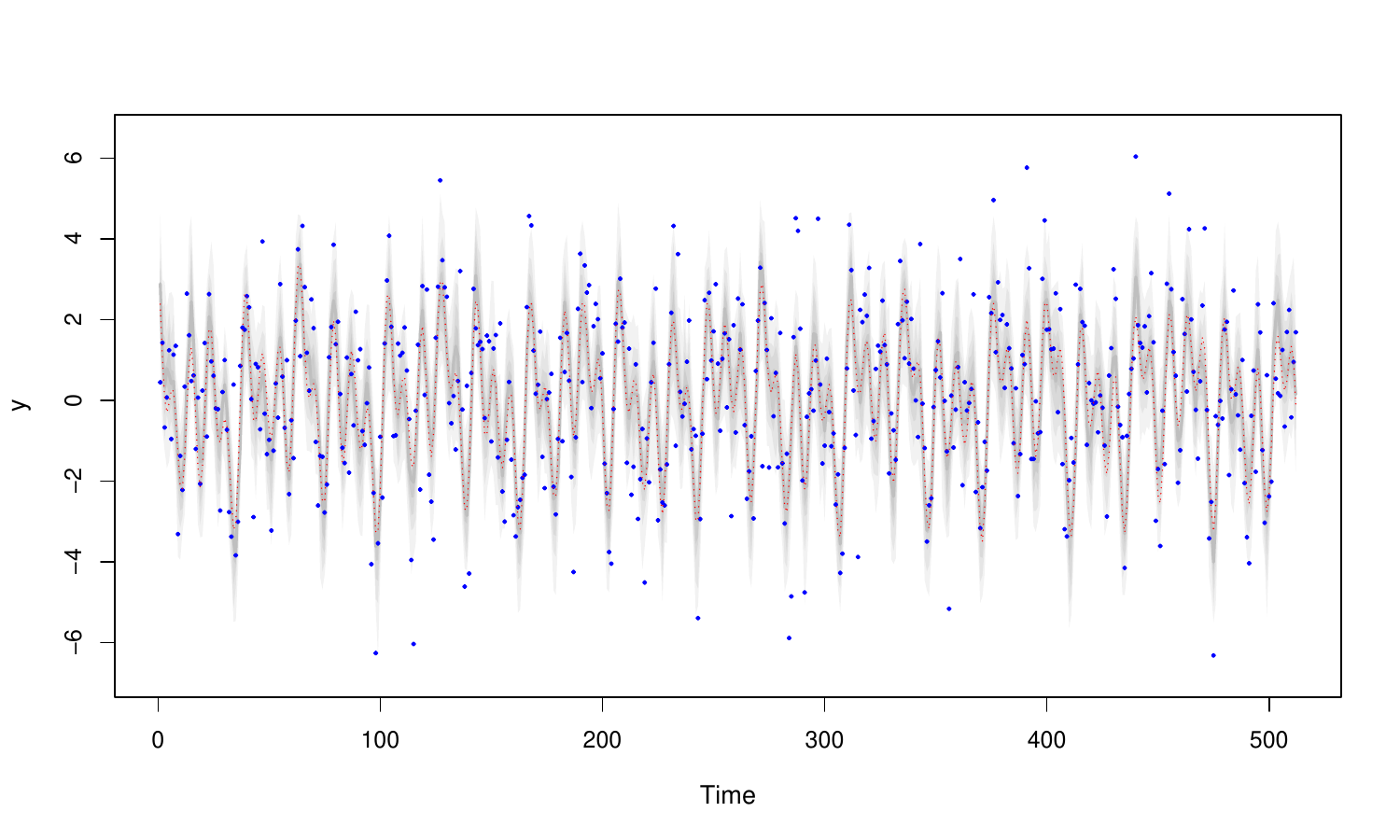} &
		\includegraphics[width=0.48\textwidth]{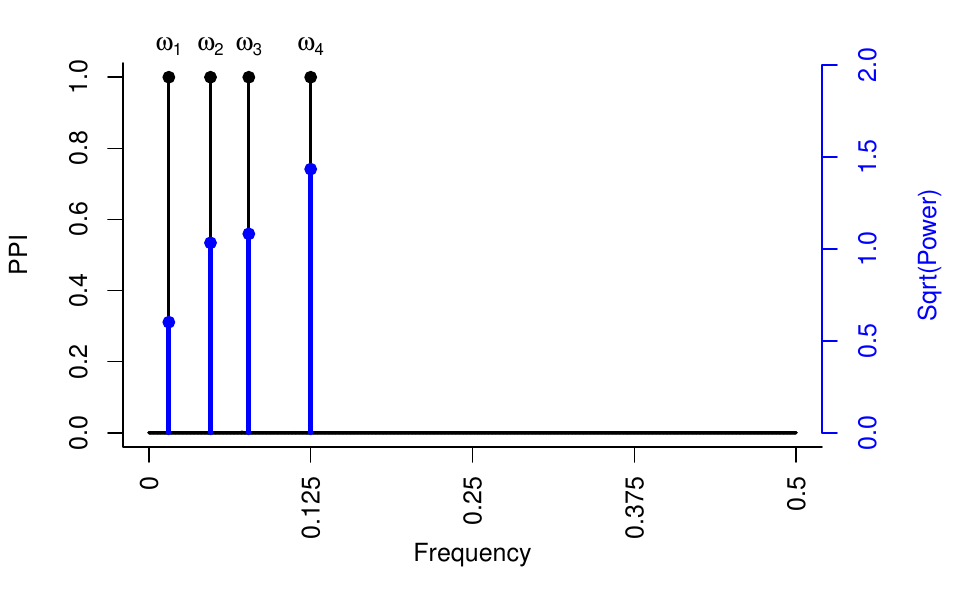}
		\\[-1.8ex]
		{\scriptsize (c)} &
		{\scriptsize (d)}
		\\[-1.2ex]
		\includegraphics[width=0.48\textwidth]{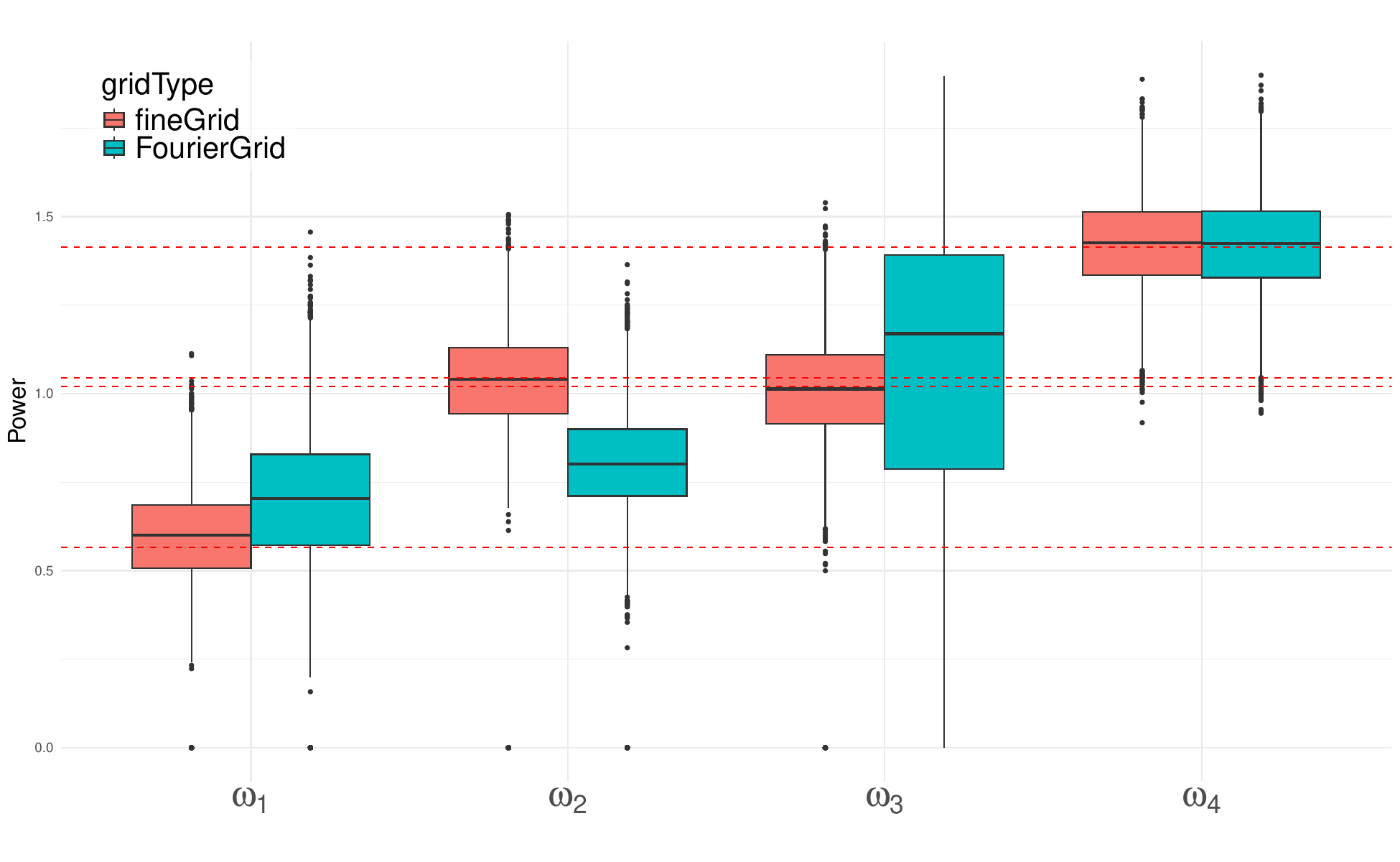} &
		\includegraphics[width=0.48\textwidth]{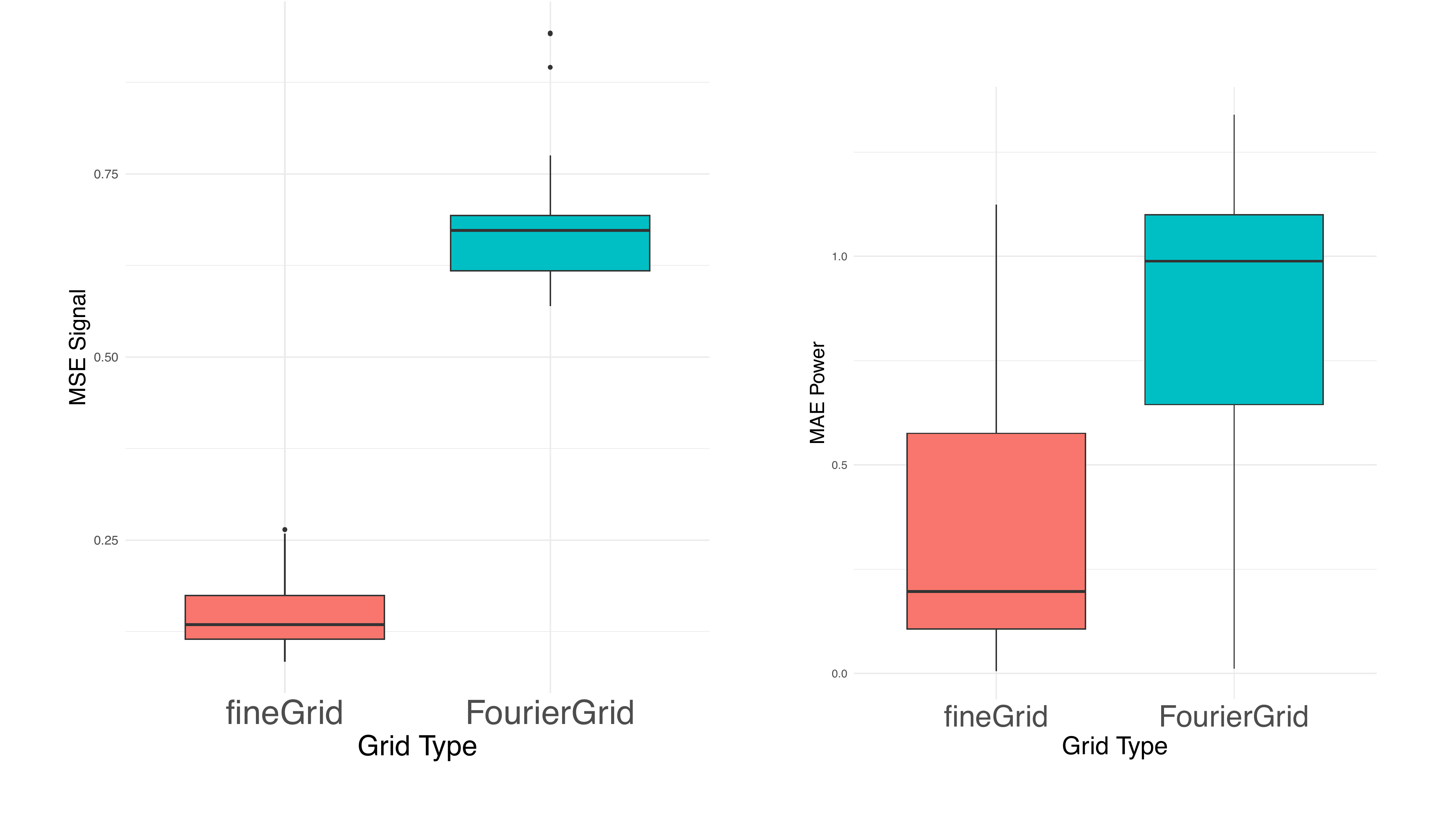}
	\end{tabular}
	\vspace{-2.2ex}

	\caption{\textbf{Simulation Study.}
		(a) Simulated data (blue dots), true signal (dotted red line), and
		100 draws from the estimated posterior predictive distribution
		(gray lines).
		(b) PPIs of the frequencies (black lines) and estimated square root
		of the power (blue lines) for frequencies with PPI $>0.5$.
		(c) Box plots of the estimated square root of power
		$\hat{P}(\omega_j)$ at each frequency, conditioned on $M=4$, with
		true power values shown as red dotted lines, for a finer grid and
		the classical Fourier grid.
		(d) Comparison of error metrics between a finer grid and the
		classical Fourier grid: MSE of the signal on the left and MAE of
		the power on the right.}
	\label{fig:data_posterior_predictive}
\end{figure}

In this simulation example, we generate a time series with $T = 512$ time points, where the periodic behavior is driven by $M = 4$ relevant frequencies $\bm{\omega} = (1/67, 1/21, 1/13, 1/8)$. The corresponding coefficients are $\beta_{11} = 0.6$, $\beta_{12} = -0.6$, $\beta_{21} = 1.0$, $\beta_{22} = 0.3$, $\beta_{31} = 1.0$, $\beta_{32} = 0.2$, $\beta_{41} = 1.0$, and $\beta_{42} = -1.0$, while the residual variance is set to $\sigma^2 = 1.5$. This selection of frequencies captures a range of oscillatory behaviors, incorporating both high-power and low-power components, as well as very low and relatively high frequencies, providing a diverse representation of the underlying signal. Figure \ref{fig:data_posterior_predictive} (a)  shows a realization of the process alongside the underlying signal. 

We set the hyperparameters for the inclusion probability $\alpha_j$, following a Beta distribution, as $a = 1$ and $b = 10$, reflecting mild prior knowledge about sparsity. We choose an uninformative prior for the residual variance, setting $\gamma_0 = \nu_0 = 0.001$, and specify the slab variance of the linear coefficients as weakly informative with $\sigma^2_\beta = 10$. When running the sampler, we selected a finely spaced grid of candidate frequencies, defined as $\omega_j = j \cdot \Delta \omega$ for $j = 0, 1, 2, \dots, L$, with a step size of $\Delta\omega = 0.0001$ and a total of $L = 5000$ frequency points.  In Section \ref{sec:finer_grid} below we show how this choice improves performances with respect to standard Fourier frequencies, which produce worse results in terms of both the accuracy of frequency estimation and the corresponding power.   We ran the MCMC sampler for 50,000 iterations, discarding the first 25,000 as burn-in. For the initial values of the periodogram, we selected the two most prominent peaks and imposed a minimum distance between selected frequencies of $d = 3$ to avoid closely spaced frequencies. Additionally, we experimented with an alternative initialization by including all frequencies in the model from the start, to assess whether the sampler could correctly identify the relevant frequencies and corresponding linear coefficients, as detailed in the Supplementary Material. Although this approach worked, initializing the sampler with only the two peaks significantly reduced the computational time required for convergence. With this parameterization, the sampler was run in approximately 50 seconds using software implemented in Julia. Convergence diagnostics for the MCMC sampler, including trace plots and the Heidelberger and Welch \citep{heidelberger1981spectral} test to assess the model's reliability, are detailed in the Supplementary Material.

The mode of the posterior probability for the number of frequencies resulted in $p(M = 4 | \bm{y}, \cdot) = 0.78$. 
Conditional on the modal number of frequencies, the estimated vector of significant frequencies were $\bm{\omega} = (0.0154, 0.0476, 0.0772, 0.125)$ and corresponding power $P(\bm{\omega}) = (0.744, 1.033, 1.082, 1.434)$. Figure \ref{fig:data_posterior_predictive} (a) displays a graphical posterior predictive check, consisting of the observations alongside 100 draws from the estimated posterior predictive \citep{gelman2007data}, showing that the model correctly captures the underlying signal. Figure \ref{fig:data_posterior_predictive} (b) illustrates the posterior inclusion probabilities (PPIs) of the frequencies together with the estimated square root of the power of the selected frequencies \( \hat{P}(\omega_j) = \sqrt{\hat{\beta}_{j1}^2 + \hat{\beta}_{j2}^2})\) for those frequencies with PPI > 0.5. Using a more stringent threshold (PPI $> 0.9$) yields the same set of selected frequencies in this simulation study. Therefore, the model correctly identifies the number of frequencies and their location, as well as the corresponding magnitude. 

\vspace{-1.5ex}

\subsubsection{Comparison Study}
\label{sec:compstudy}
We evaluate the performance of our discrete spike-and-slab approach in identifying relevant frequencies against regularization techniques that operate on the same set of Fourier frequencies. Specifically, our methodology is benchmarked against four alternative approaches, each representing a different strategy for frequency selection in time series analysis: (i) \textbf{spectralLASSO}: This approach adopts the same saturated regression model framework as our proposed method. However, instead of using a spike-and-slab prior for frequency selection, variable selection is achieved via LASSO penalization \citep{tibshirani1996regression}. This penalization method imposes an $\ell_1$-norm constraint on the linear coefficients, which promotes sparsity by shrinking many of the Fourier coefficients to zero. 
(ii) \textbf{spectralEN} (spectral Elastic Net): Like LASSO, Elastic Net \citep{zou2005regularization}  begins with a saturated regression model over all Fourier frequencies. However, it applies both $\ell_1$- and $\ell_2$-norm constraints. While this dual-penalty structure can be advantageous in certain applications, the Elastic Net approach may include irrelevant frequencies due to its tendency to retain correlated predictors, potentially affecting model interpretability and precision in frequency selection. 
(iii) \textbf{spectralSCAD}: This method replaces the LASSO penalty with the Smoothly Clipped Absolute Deviation (SCAD, \citealt{fan2001variable}) penalty, a non-convex regularization technique known to reduce shrinkage bias for large coefficients while maintaining sparsity. SCAD can offer improved frequency selection when the true signal is sparse and the Fourier basis is highly correlated. 
(iv) \textbf{spectralMCP}: This approach employs the Minimax Concave Penalty (MCP, \citealt{zhang2010nearly}), another non-convex alternative designed to balance sparsity and estimation bias. MCP aggressively shrinks small coefficients while preserving larger ones, often leading to more precise frequency recovery under low signal-to-noise scenarios. 
We implemented spectralLASSO, spectralEN, spectralSCAD, and spectralMCP ourselves, using the \texttt{glmnet} and \texttt{ncvreg} packages in \texttt{R}.
We utilized the following measures to assess accuracy in estimating both frequency locations and spectral power: (i) Mean Squared Error of the signal, calculated as $MSE_S = \frac{1}{T}\sum_{t=1}^{T}(f_t - \hat{f}_t)^2$,
(ii) Absolute Error of the power, $AE_P = \left| \sum_{j=1}^{M} P(\omega_j) - \sum_{j=1}^{\hat{M}} \hat{P}(\omega_j)\right|$, and (iii) Absolute Error of the frequencies, $AE_F = \left| \sum_{j=1}^{M} \omega_j - \sum_{j=1}^{\hat{M}} \hat{\omega}_j\right|$, where the hat notation denotes estimated parameters.  Here, $f$ and $\hat{f}$ represent the true and estimated signals, respectively (e.g., $f_t = \sum_{j=1}^{M} \left\{ \beta_{j1} \cos(2\pi\omega_j t) + \beta_{j2} \sin(2\pi\omega_j t) \right\}$).

\begin{figure}[ht!]
	\centering
	\includegraphics[width=1\textwidth]{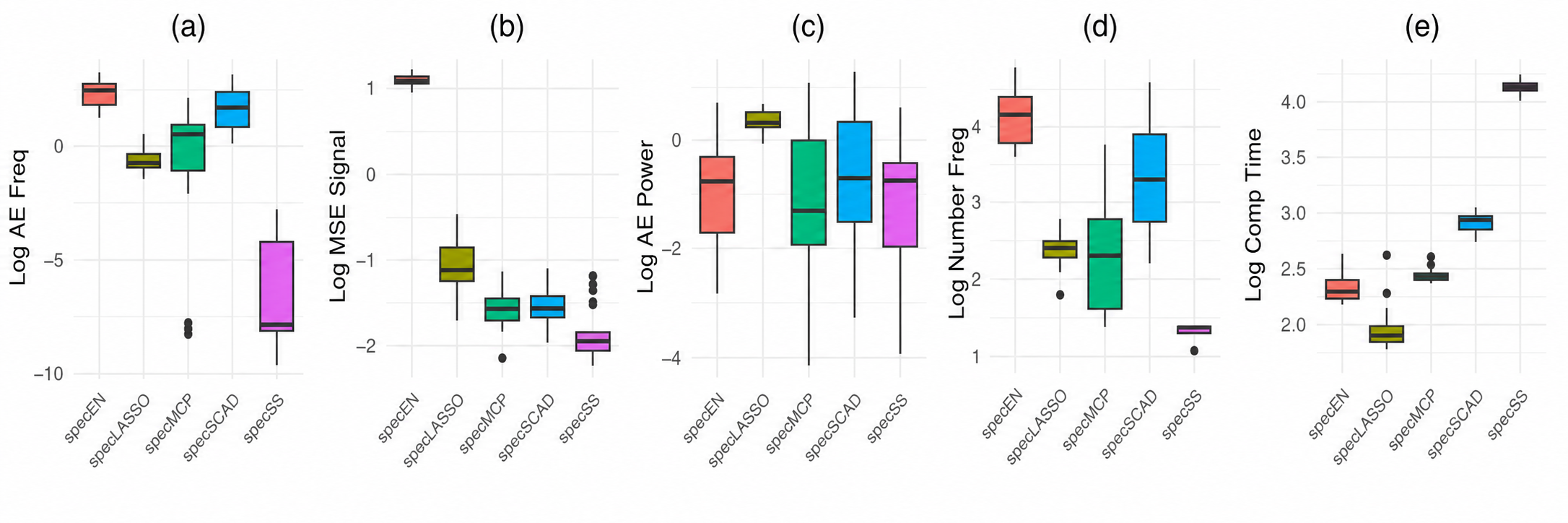}
	\caption{\textbf{Simulation Study.} Comparison of the performance of our proposed method (spectralSS) across 50 simulations against spectralLASSO, spectralEN, spectralSCAD and spectralMCP. Boxplots illustrate (a) $\log(AE_F)$, (b) $\log(MSE_S)$, (c) $\log(AE_P)$, (d) the logarithm of the selected number of frequencies, and (e) log computational time (in seconds). }
	\label{fig:boxplot_methods}
\end{figure}

The performance of our proposed approach across 50 simulations, compared with spectralLASSO, spectralEN, spectralSCAD, and spectralMCP, is illustrated in Figure~\ref{fig:boxplot_methods}. We present boxplots of (a) $\log(AE_F)$, (b) $\log(MSE_S)$, (c) $\log(AE_P)$, (d) the logarithm of the selected number of frequencies, and (e) log computational time. Our proposed method, \textbf{spectralSS}, demonstrates superior accuracy across all performance metrics. In particular, it achieves the lowest median error in frequency estimation, signal reconstruction, and power spectrum approximation, while also recovering the true number of frequencies more consistently than the other approaches. Although spectralSS is computationally more intensive than the regularization-based methods, the trade-off results in significantly improved performance. Among the penalization-based competitors, the non-convex methods—spectralSCAD and spectralMCP—generally outperform spectralLASSO and spectralEN in terms of accuracy, but tend to be less sparse or less stable in frequency selection than spectralSS. Notably, spectralEN exhibits a tendency to overestimate the number of active frequencies. This scenario considers a high-dimensional setting with a very small sample size relative to the number of candidate frequencies; specifically, the saturated spectral design matrix has dimension $512 \times 10{,}000$. Despite this challenging regime, our approach shows remarkable power and precision, consistently identifying the relevant components with minimal estimation error.

\subsubsection{Finer Grid of Frequencies} \label{sec:finer_grid}

Next, we investigate the advantages of selecting a finer grid of frequencies in comparison to the classical Fourier grid. To explore this, we simulated 50 time series based on the same simulation scenario described above, using both types of frequency grids. Figure \ref{fig:data_posterior_predictive} (c) presents box plots of the estimated square root of the power $\hat{P}(\omega_j)$ at each frequency, for each frequency grid, with the results conditioned on the estimated number of frequencies $M = 4$. The dotted horizontal red lines correspond to the true values of the power. For this example, selecting a finer frequency grid generally outperforms the Fourier grid, except for the frequency 0.125 (i.e., periodicity 8) where the results are almost identical.  Figure\ref{fig:data_posterior_predictive} (d) displays box plots of the $MSE_S$ of the signal and the $AE_P$ for both methods. Thus, the finer frequency grid consistently outperforms the classical Fourier grid in terms of both signal reconstruction and power estimation. 

In the Supplementary Material we further investigate the behavior of the proposed method in scenarios where oscillatory components are closely spaced in the frequency domain and compare its performance with alternative methods. We also examine the effectiveness of our proposed method in detecting spectral peaks when there is a mismatch between the model and the data-generating process. Finally, we present results from a sensitivity analysis.

\subsection{Bivariate Time Series} \label{sec:bivariate_simul}

\begin{figure}
	\centering
	\begin{minipage}[t]{0.245\textwidth}
		\centering
		{\scriptsize (a)}\\[-0.5ex]
		\includegraphics[width=\linewidth]{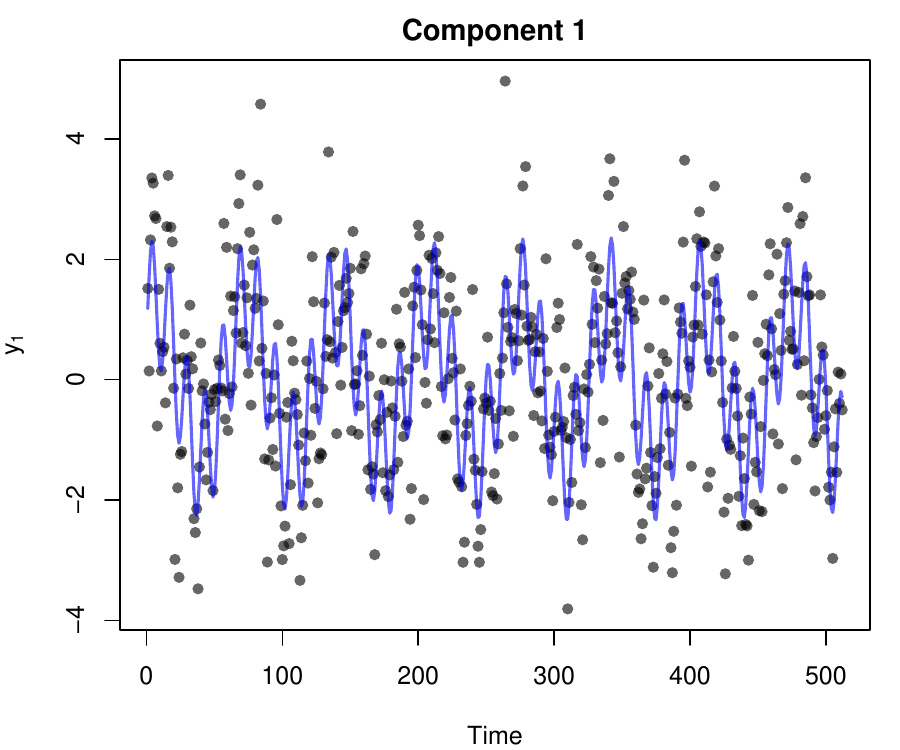}
	\end{minipage}\hspace{-0.6ex}%
	\begin{minipage}[t]{0.245\textwidth}
		\centering
		{\scriptsize (b)}\\[-0.5ex]
		\includegraphics[width=\linewidth]{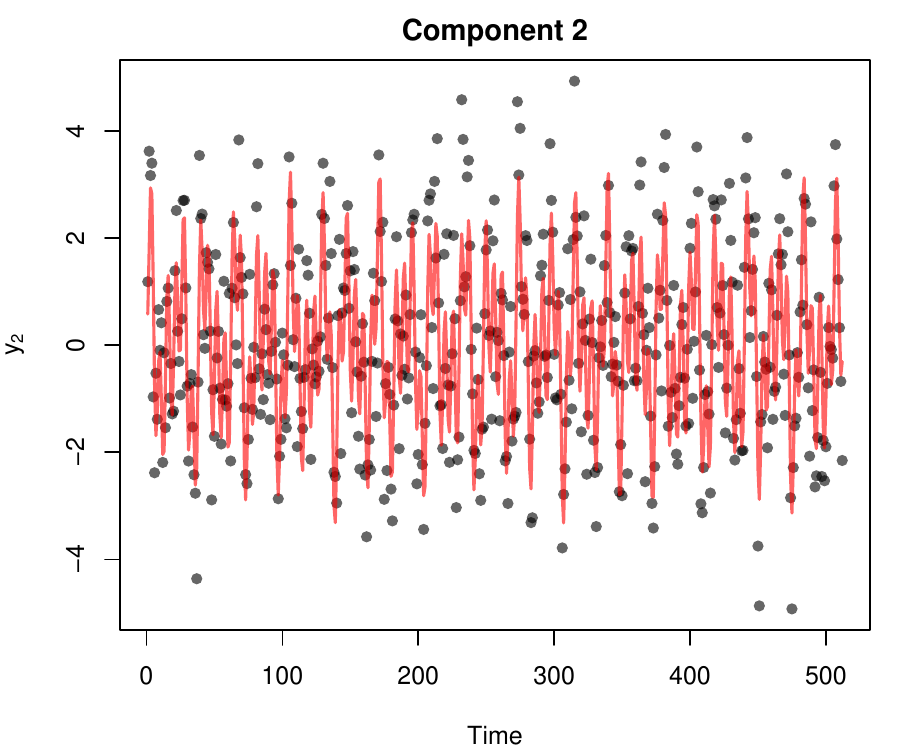}
	\end{minipage}\hspace{-0.6ex}%
	\begin{minipage}[t]{0.245\textwidth}
		\centering
		{\scriptsize (c)}\\[-1.2ex]
		\includegraphics[width=\linewidth]{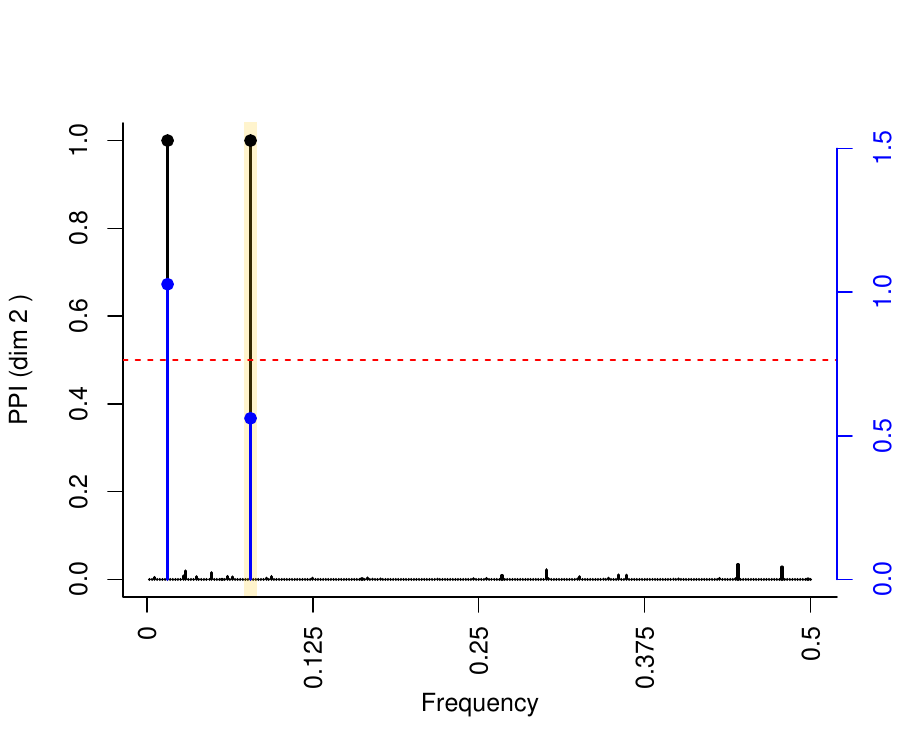}
	\end{minipage}\hspace{-0.6ex}%
	\begin{minipage}[t]{0.245\textwidth}
		\centering
		{\scriptsize (d)}\\[-1.2ex]
		\includegraphics[width=\linewidth]{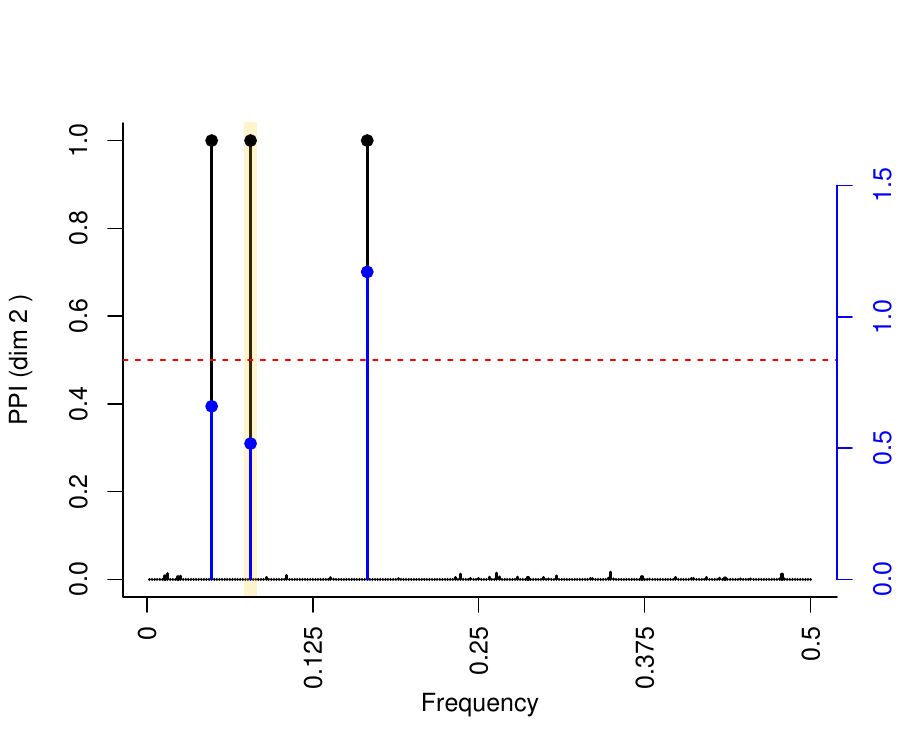}
	\end{minipage}

	\vspace{-1.5ex}

	\caption{\textbf{Simulation Study.}
		(a)--(b) Simulated observations (dots) and true underlying signals
		(lines) for the two components.
		(c)--(d) Posterior inclusion probabilities (PPIs; black lines) and
		estimated square roots of signal power (blue lines) for frequencies
		with PPI greater than 0.5.
		The shaded region highlights a frequency that is jointly selected
		across both components but with distinct amplitudes, illustrating the
		model’s ability to recover partially shared spectral features.}
	\label{fig:multiv_simul}
\end{figure}

Next, we consider a bivariate simulation with \( T = 512 \) time points, where each component exhibits distinct but partially overlapping periodic behavior. Specifically, the first series is generated using \( M_1 = 2 \) active frequencies, \( \bm{\omega}_1 = (1/67,\ 1/13) \), while the second relies on \( M_2 = 3 \) active frequencies, \( \bm{\omega}_2 = (1/21,\ 1/13,\ 1/6) \). The corresponding Fourier coefficients are set to \( \bm{\beta}_{11} = (1.0,\ -0.4) \), \( \bm{\beta}_{21} = (0.8,\ 1.0) \) for the first dimension, and \( \bm{\beta}_{12} = (0.9,\ 0.5) \), \( \bm{\beta}_{22} = (-1.0,\ 0.8) \), \( \bm{\beta}_{32} = (0.7,\ -0.9) \) for the second. Independent Gaussian noise is added to each series, with component-specific variances set to \( \sigma_1^2 = 1 \) and \( \sigma_2^2 = 1 \).
Notably, the two components share a common frequency at \( 1/13 \), although their amplitudes and phases differ, yielding both shared and dimension-specific structure. We specified the Dirichlet hyperparameter for $\bm{\pi}$ as \( \boldsymbol{\alpha} = (10.0,\ 3.0,\ 3.0,\ 3.0) \), leading to a prior that favors configurations with fewer jointly selected frequencies across components. Figure \ref{fig:multiv_simul} (a)  shows a realization of the process alongside the underlying signal. 

The posterior mode for the number of active frequencies was \( M_1 = 2 \) and \( M_2 = 3 \) for the two components, with respective posterior probabilities \( p(M_1 = 2 \mid \mathbf{y}, \cdot) = 0.87 \) and \( p(M_2 = 3 \mid \mathbf{y}, \cdot) = 0.82 \).  
Conditional on these modal values, the estimated sets of dominant frequencies were \( \boldsymbol{\omega}_1 = (0.016, 0.076) \) and \( \boldsymbol{\omega}_2 = (0.047, 0.076, 0.166) \), with corresponding estimated signal power \( P(\boldsymbol{\omega}_1) = (1.15, 0.76) \) and \( P(\boldsymbol{\omega}_2) = (0.90, 0.61, 1.07) \).
Figure~\ref{fig:multiv_simul}(b) displays the PPIs across the candidate frequency grid for each component, along with the estimated square root of the signal power for frequencies with \(\mathrm{PPI} > 0.5\). The model successfully recovers the true number and location of the generating frequencies in both series and, crucially, identifies a shared frequency at \( \omega = 0.076 \) that is included in both components but with distinct amplitudes—highlighting its ability to capture partially overlapping yet component-specific spectral structure.
In the Supplementary Materials we further show that similar results are obtained when the data are generated with correlated errors across components, despite the independence assumption in the fitted model.

\vspace{-1ex}

\section{Applications to Wearable Device Data}
\label{sec:Data}
Wearable biosensors offer an unprecedented opportunity to non-invasively monitor physiological rhythms in real-world settings \citep{de2024state}. We apply our proposed framework on two distinct datasets: (i) actigraphy data from epilepsy patients, and (ii) temperature-activity data from a healthy subject.

\subsection{Wearable Activity Monitoring in Epilepsy Studies} 
\label{sec:application}
Actigraphy data provide a non-invasive means of tracking rest-activity cycles, capturing key metrics such as sleep duration and daytime movement levels. In epilepsy research, these data can reveal patterns associated with seizure frequency and psychiatric symptoms like depression and anxiety \citep{abboud2023actigraphic}. Identifying the specific frequencies, amplitudes, and phases within these cycles presents a challenge, yet these components may hold promise as a biomarkers that may encode states of neurological or psychiatric compromise \citep{smagula2024sleep}. 

\subsubsection{Study Framework and Data Preprocessing}

Our analysis focused on data previously introduced by \citet{abboud2023actigraphic}. Participants were enrolled through the Baylor Comprehensive Epilepsy Clinic in Houston, TX. The study targeted adults aged 18 to 75 with partial-onset seizures. Each participant wore an FDA-approved Actiwatch-2 device (Philips Respironics), highly durable and widely trusted in clinical research, continuously on their non-dominant wrist, including during sleep and showers. The Actiwatch-2 uses piezoelectric sensors calibrated to measure wrist movement via a single-axis accelerometer. For this analysis, we considered six individuals.
We conducted a series of preprocessing steps on the raw activity data, adapting established methods \citep{huang2018hidden, abboud2023actigraphic}. First, we applied a logarithmic transformation (adding one to each value). Next, we used a moving average filter with a 15-point window, equivalent to a 15-minute span, followed by downsampling by averaging every five consecutive points, yielding a 5-minute sampling rate. This process resulted in time series with lengths $T$ ranging from 1725 to 2301 data points. 

\subsubsection{Parameter Settings}

For our proposed methodology, we selected a grid of Fourier frequencies, ensuring the grid was sufficiently dense given the length of the time series. We set \( a = 1 \) and \( b = 10{,}000 \) to promote a high degree of prior sparsity for the frequencies. This choice is guided by the sensitivity analysis presented in the Supplementary Material, where we evaluated the effect of varying \( b \) on frequency recovery across a wide range of sample sizes. Notably, for time series lengths comparable to those encountered in the real data applications, the setting \( b = 10{,}000 \) provided a favorable balance—yielding low error in frequency selection and signal reconstruction, while closely approximating the true number of active frequencies. While alternative strategies such as empirical Bayes via MMLE or tuning via cross-validation could be explored, our empirical results indicate that the chosen value of \( b \) offers robust and competitive performance in the considered regimes. Additionally, we imposed a minimum frequency distance of \( d = 5 \) to ensure sufficient separation between selected components. As discussed in Section \ref{sec:selectin_minimum_d}, this constraint is particularly relevant in biological time series, such as actigraphy, where dominant rhythms like circadian ($\sim$ 24 hours), ultradian (e.g., 12 hours), and infradian cycles are of interest. On a fine frequency grid, nearby bins often correspond to periods that are nearly indistinguishable in physiological terms (e.g., 24.1 vs 23.7 hours). Setting \( d = 5 \) reduces the risk of selecting redundant or collinear frequencies in such dense regions, enhancing interpretability while stabilizing inference. The MCMC sampler was run for $50,000$ iterations, with the first $25,000$ discarded as burn-in. Posterior predictive graphical checks, presented in the Supplementary Material, demonstrate that the estimated model accurately captures the structure of the observed data.




\subsubsection{Results}

\begin{figure}[ht!]
	\centering
	\includegraphics[width=1.0\textwidth]{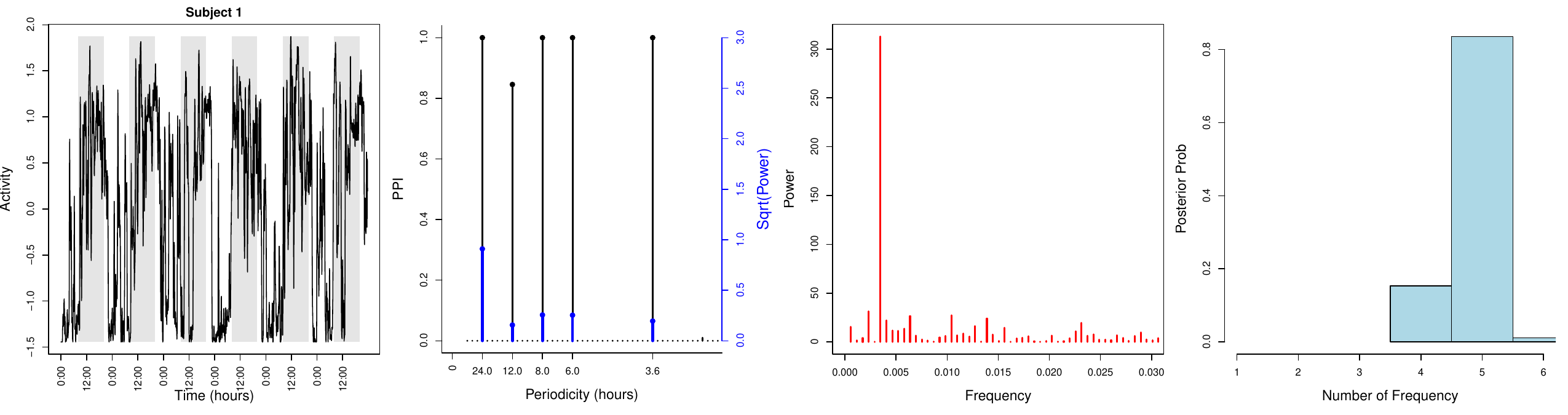}
\caption{\textbf{Wearable Activity Monitoring.} Left panel: standardized activity time series, with gray bands indicating 8 AM--8 PM; second panel: estimated periodicities with posterior inclusion probabilities and square-root spectral power; third: raw periodogram; and fourth: posterior distribution of the number of selected frequencies.}
	\label{fig:results_data_subject1}
\end{figure}

Figure \ref{fig:results_data_subject1} presents the results for Subject 1, while the corresponding results for the remaining five subjects are shown in the Supplementary Material. The first panel displays the observed standardized actigraphy time series. The second panel shows the estimated frequencies alongside their PPIs and corresponding square-root spectral power, with the raw periodogram in the third panel serving as a baseline for comparison. The fourth panel illustrates the posterior distribution over the number of selected frequencies. These results highlight the efficacy of the proposed spike-and-slab prior in identifying the most significant frequencies that contribute to the observed oscillatory patterns. The number of selected frequencies varies across individuals, reflecting differences in the periodic structure of their actigraphy data. Some subjects exhibit relatively simple oscillatory patterns characterized by a few dominant frequencies, while others display more intricate dynamics with a greater number of selected periodic components. As expected, all subjects exhibited a strong circadian (24-hour) periodicity, indicating the robust presence of diurnal rest-activity cycles. Subjects 3, 4, and 5, shown in the Supplementary Material, share similar oscillatory patterns, with ultradian (less than 24 hours) oscillations observed at approximately 12 and 6 hours. Subject 6 exhibits a comparatively simpler oscillatory pattern, characterized primarily by periodicities at approximately 24 and 12 hours. Notably, the periodogram for Subject 4 appears noisier compared to the others. In contrast, Subject 2 demonstrates an additional periodicity at approximately 13 hours, while Subject 1 presents a more complex oscillatory pattern. This is evidenced by the greater number of selected frequencies, which include ultradian oscillations at approximately 12, 8, 6, and 3.6 hours, in addition to the dominant circadian rhythm.

Our findings hold significant implications for monitoring circadian and ultradian rhythms in clinical populations. By automatically identifying key periodic components, the proposed methodology could serve as a powerful tool for tracking variations in rest-activity rhythms, potentially offering new insights into the relationship between these rhythms and neurological states, such as seizure frequency and psychiatric symptoms, including depression or anxiety. Indeed, the connection between mental health and actigraphy has primarily focused on circadian rhythms, while interindividual variability in the expression of ultradian rhythms has not been extensively explored. This is because traditional periodography often struggles to distinguish genuine ultradian rhythms from artifactual periodicities, as the dominance of the circadian oscillator can obscure ultradian rhythmicity.  Furthermore, 
exogenous factors, such as a regular 8-hour eating schedule, can also produce pronounced 8-hour ultradian rhythms. Recent findings \citep{adhyapak2024stability} reveal unique signatures of ultradian rhythms expression between individuals, with these signatures also exhibiting temporal variability.

We further compared our proposed approach to spectralLASSO and spectralEN. SpectralLASSO identified an average of 322 frequencies across the six subjects, whereas spectralEN identified an average of 353 frequencies. This indicates that both methodologies struggle to automatically isolate the dominant frequencies from the broader set of candidates, with spectralEN tending to select a larger number of frequencies than spectralLASSO. These comparisons highlight the strengths of the spike-and-slab prior, which effectively balances sparsity and model fidelity. By doing so, it enables the identification of a parsimonious set of frequencies that closely align with prominent peaks in the periodogram, while also mitigating the risk of overfitting to noise. Additionally, the posterior probability intervals provide a clear and interpretable measure of uncertainty in frequency estimation. These findings underscore the advantages of selection approaches over traditional spectral analysis methods, as they automate the selection of meaningful frequencies in a principled manner.


\subsection{Bivariate Case Study}
\begin{figure}[htbp]
	\centering
	\begin{minipage}[t]{0.48\textwidth}
		\centering
		{\scriptsize (a)}\\[-0.5ex]
		\includegraphics[width=\linewidth]{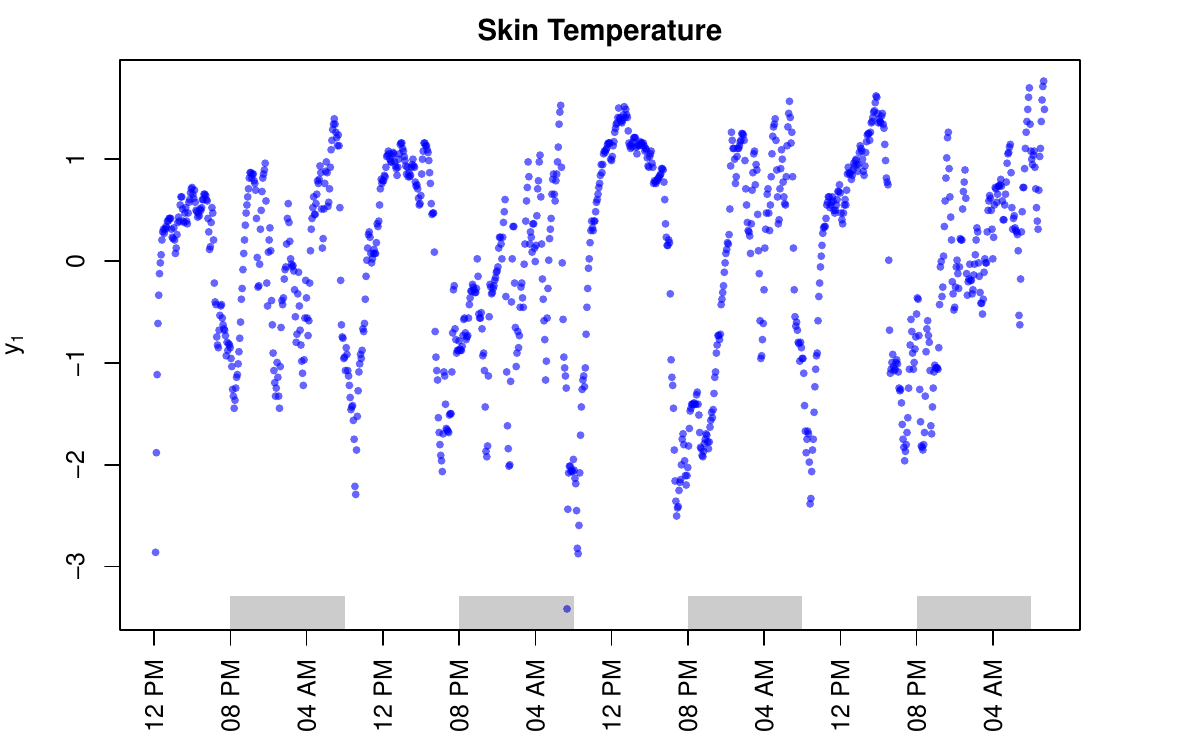}
	\end{minipage}\hspace{0.4em}%
	\begin{minipage}[t]{0.48\textwidth}
		\centering
		{\scriptsize (b)}\\[-0.5ex]
		\includegraphics[width=\linewidth]{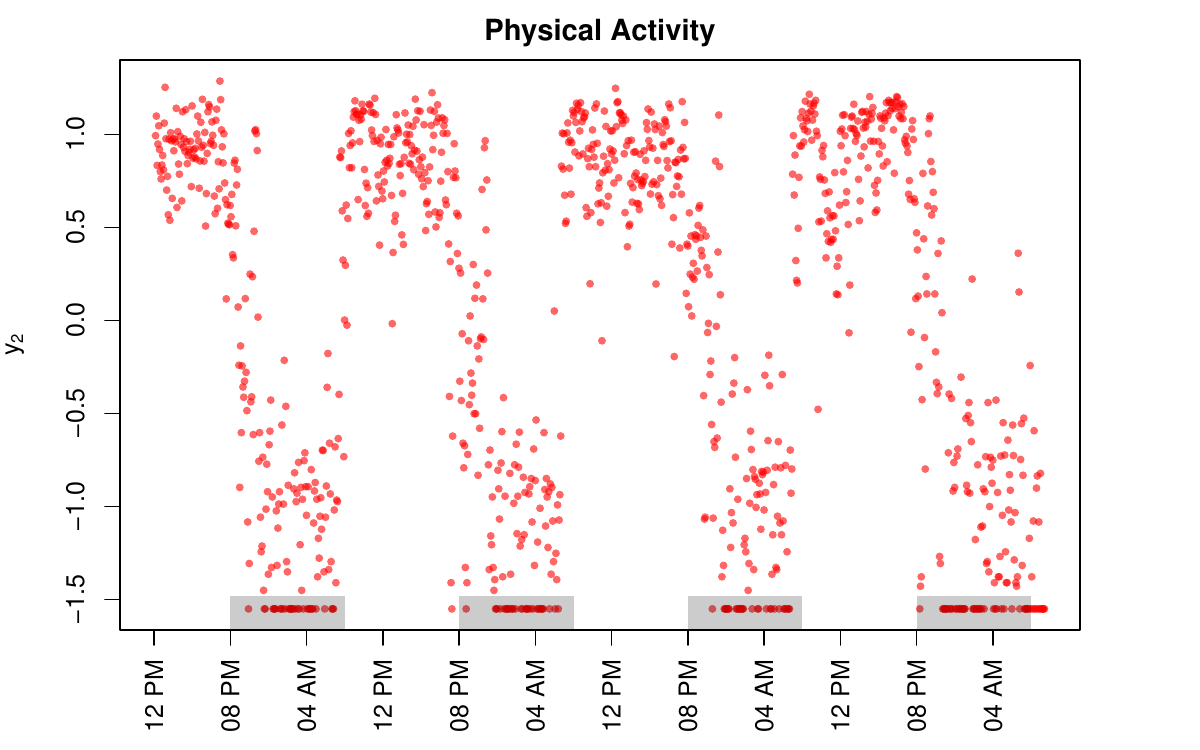}
	\end{minipage}

	\vspace{-0.2ex}

	\begin{minipage}[t]{0.48\textwidth}
		\centering
		{\scriptsize (c)}\\[-1.0ex]
		\includegraphics[width=\linewidth]{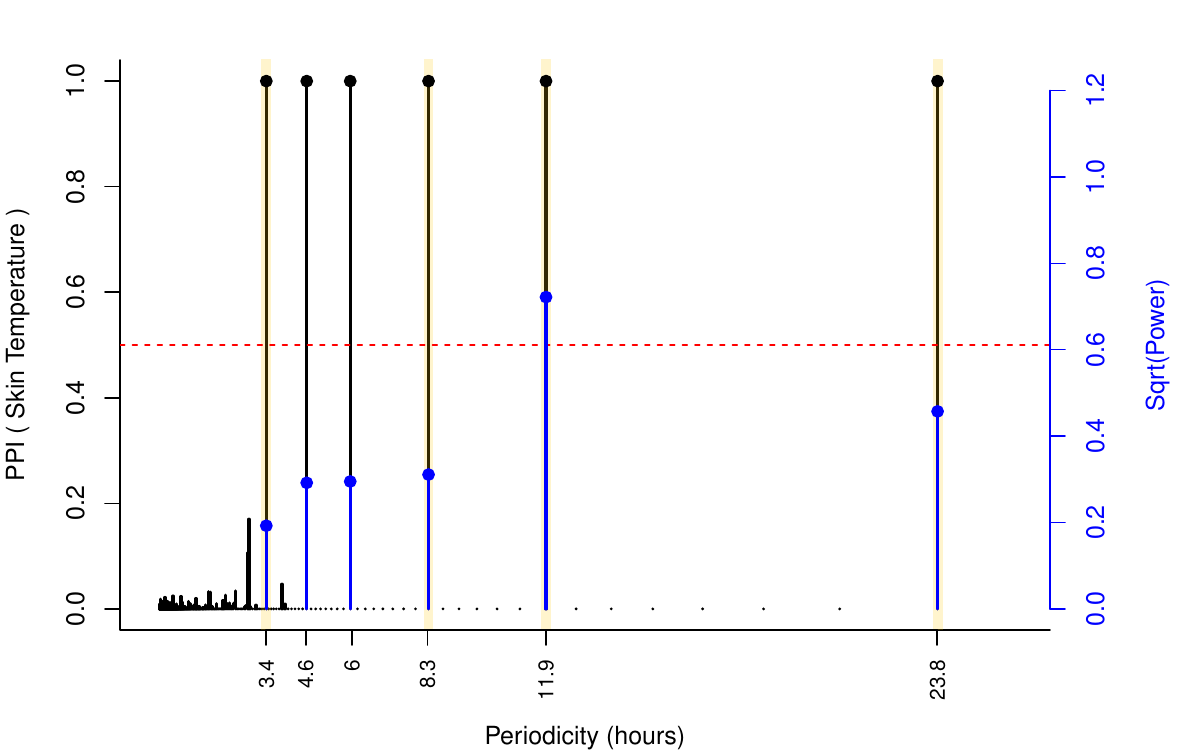}
	\end{minipage}\hspace{0.4em}%
	\begin{minipage}[t]{0.48\textwidth}
		\centering
		{\scriptsize (d)}\\[-1.0ex]
		\includegraphics[width=\linewidth]{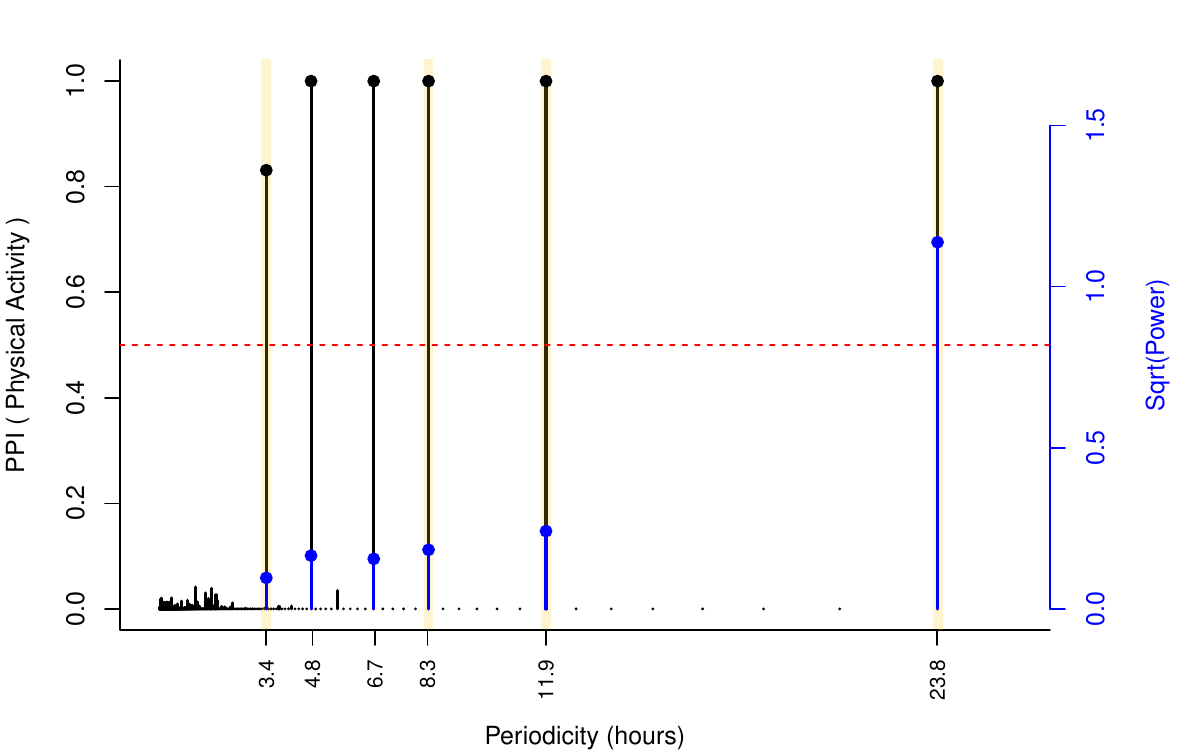}
	\end{minipage}

	\vspace{-1ex}

	\caption{\textbf{Bivariate Case Study.}
		(a)--(b) Time series of skin temperature and physical activity, with
		nighttime intervals shaded in gray from 8 PM to 8 AM.
		(c)--(d) PPIs (black) and estimated power (blue) across periodic
		components, with shared periodicities highlighted in orange.}
	\label{fig:multiv_appl}
\end{figure}

Skin temperature and physical activity are two circadian-regulated signals with known physiological coupling. While core body temperature usually peaks in the late afternoon, skin temperature at the body’s surface tends to rise during the early night, helping the body release heat and fall asleep more easily. 
This antiphase dynamic reflects the underlying circadian drive on thermoregulation and behavior, and has been leveraged in personalized medicine approaches such as chronotherapy. 
In recent years, mobile health platforms have enabled simultaneous and continuous telemetric monitoring of skin surface temperature and physical activity, offering a unique opportunity to study these rhythms.
The data analyzed here were originally collected using the PiCADo mobile e-health platform and have been studied separately in previous works. Under a stationarity assumption, \citet{komarzynski2018picado} analyzed the skin temperature time series and identified both strong 12-hour and 24-hour rhythms. \citet{hadj2020bayesian} applied a nonstationary oscillatory model with an unknown change point to the same data, demonstrating that skin temperature alone can be used to detect sleep periods and reveal ultradian oscillations during the night—oscillations that are consistent with transitions between REM and non-REM sleep stages. \citet{huang2018hidden} modeled the activity data using a circadian hidden Markov model, while \citet{hadj2023bayesian} employed a semi-Markov switching process for activity dynamics. To our knowledge, this is the first time that temperature and activity data from this dataset have been jointly analyzed within a unified statistical framework. 
Our preprocessing procedure follows \citet{komarzynski2018picado}, applying a square-root transformation to the raw activity counts and subsequently standardizing both time series. The resulting bivariate signals, spanning four consecutive days, are displayed in Figure~\ref{fig:multiv_appl}(a).

\subsubsection{Results}
We adopt the same hyperparameter settings of the univariate application (Section~\ref{sec:application}) and specify the Dirichlet prior parameter $\bm{\alpha}$ for the inclusion patterns as in the bivariate simulation study (Section~\ref{sec:bivariate_simul}). The posterior distribution over the number of selected frequencies indicates that physical activity most likely contains six dominant periodic components, with posterior mode \( P(M_1 = 6) = 0.35 \), followed by seven components with \( P(M_1 = 7) = 0.29 \). In contrast, skin temperature  shows a posterior mode at seven components (\( P(M_2 = 7) = 0.33 \)), with substantial probability mass at six and eight components (\( P(M_2 = 6) = 0.23 \), \( P(M_2 = 8) = 0.23 \)). Posterior predictive checks indicate that the model faithfully reproduces the temporal dynamics of both signals, as evidenced by close alignment between observed data and posterior predictive draws (see Supplementary Material).

Figure~\ref{fig:multiv_appl}(b) displays PPIs and estimated signal strength for the periodic components of physical activity (right panel) and skin temperature (left panel). In the activity signal, a dominant circadian rhythm emerges at approximately 24 hours. In contrast, while the temperature signal also shows evidence of a circadian component, its posterior amplitude is notably lower, suggesting reduced circadian influence relative to activity. Both modalities reveal rich ultradian dynamics. In particular, common components are identified near 3.4 and 8.3 hours, although their relative contribution differs between signals. For instance, the 8.3-hour oscillation exhibits high amplitude in temperature but is comparatively weaker in activity, whereas the 3.4-hour component, though less prominent, shows similar magnitude across both modalities. The presence of a shared ultradian rhythm near 3.4 hours is biologically meaningful, as it likely reflects coupled regulation between physical activity and thermoregulatory systems \citep{goh2019episodic}. In contrast, temperature shows additional specific components—such as at 6 hours and 11.9 hours—that are absent or significantly smaller in the activity signal, suggesting internal thermoregulatory cycles that operate independently of physical movement. The 11.9-hour component, consistent with a circahemidian rhythm, may emerge as a subharmonic of the underlying circadian oscillator \citep{grant2018evidence}. Moreover, some ultradian components appear modality-specific, including the 4.6-hour and 6-hour rhythms in temperature and the 4.8-hour and 6.7-hour components in activity, underscoring partially independent rhythmic processes.
Overall, our analysis reveals that temperature and activity signals follow different, yet partly overlapping, rhythmic patterns. While physical activity primarily captures the circadian cycle, skin temperature exhibits a broader range of ultradian periodicities, emerging as a complementary biomarker for chronophysiological dynamics. The bivariate framework offers a flexible model of how physiological rhythms work together or differ. 

\section{Discussion} \label{sec:discussion}

We have introduced a Bayesian spike-and-slab framework for spectral analysis, offering a robust and interpretable approach to identifying meaningful periodicities in time series data. The proposed approach integrates frequency selection and dimensionality reduction using a refined grid of candidate frequencies, achieving high-resolution recovery of oscillatory components while enforcing sparsity through a spike-and-slab prior. We have designed an efficient stochastic search algorithm to explore the posterior space and estimate posterior inclusion probabilities to assess the significance of each frequency. Using simulated data, we have shown superior performance of the proposed method in identifying relevant frequencies and estimating spectral power compared to alternative approaches. Additionally, we have shown that using a finer frequency grid enhances accuracy in estimating power and reduces error metrics relative to the classical Fourier grid, highlighting the advantages of higher-resolution frequency selection. An application to actigraphy data from individuals with partial-onset seizures demonstrated the model's ability to identify clinically relevant rhythms, highlighting its potential as a tool for linking rest--activity patterns to neurological and psychiatric states. We have further extended the framework to the multivariate setting through a hierarchical prior on frequency inclusion, enabling the identification of both shared and component-specific rhythms across multiple signals. Applied to bivariate wearable measurements from a healthy individual to jointly model wrist activity and skin temperature, the model successfully captured partially overlapping ultradian components and known physiological coupling between thermoregulatory and behavioral rhythms. These results underscore the flexibility of the proposed method for multichannel physiological data and support its use in future applications of chronobiological modeling and digital health analytics.

Extensions in several directions may further broaden the utility and robustness of the proposed framework. First, the current model introduces a minimum separation parameter $d$ that enforces a spacing constraint between selected frequencies through the indicator function $I(|\omega_j - \omega_k| \ge d)$. This parameter acts as a resolution control that prevents the model from selecting clusters of nearly indistinguishable frequencies on a refined grid. An interesting extension would be to treat $d$ as a random parameter and allow the data to inform the degree of separation enforced by the model. However, because the constraint directly determines the set of admissible configurations of the inclusion indicators, different values of $d$ correspond to different model spaces. In particular, the support of the prior on the inclusion indicators depends on $d$, and the constrained prior distribution involves a normalization constant equal to the number of admissible frequency configurations satisfying the separation constraint. This quantity depends on $d$ and is combinatorial in nature, which would substantially complicate posterior computation and the design of the MCMC algorithm. For this reason, in the present work $d$ is treated as a user-specified resolution parameter. Nevertheless, allowing the degree of frequency separation to be learned from the data is an interesting research direction.

A second limitation concerns scalability in the multivariate extension. In the proposed formulation, each candidate frequency $\omega_j$ is associated with a joint inclusion pattern $(z_j^{(1)},\ldots,z_j^{(D)})$ across the $D$ components, leading to $2^D$ possible configurations. As a result, the number of possible joint patterns grows exponentially with the dimension $D$, increasing both the number of parameters in the prior specification and the computational cost of exploring the corresponding model space through MCMC. While this formulation allows the model to capture shared and component-specific oscillatory behavior in a coherent probabilistic framework, the exponential growth in the number of configurations may limit scalability when the number of components becomes large. In practice, the approach remains computationally feasible for small to moderate values of $D$, such as the bivariate case considered in our experiments, where the joint inclusion structure provides a flexible mechanism for modeling partially shared rhythms across signals. More scalable formulations for higher-dimensional multivariate time series, potentially through more parsimonious joint priors or structured dependence across components, is another promising direction for future work.

Additional extensions could enhance the modeling framework. Alternative basis functions, such as wavelets or splines, could improve performance in the presence of nonstationary or localized temporal features. More flexible error structures, such as autoregressive or state-space models, would allow to better capture residual temporal dependence when needed. From a theoretical standpoint, posterior contraction rates and model selection consistency under spike-and-slab priors in the spectral domain could provide valuable insights into the estimator’s properties. Additional gains could come from integrating time-varying covariates or subject-level information, for a deeper exploration of how rhythmic dynamics relate to external influences. Finally, improvements in computational efficiency via parallelization or variational inference would support applications to larger datasets. Although developed for wearable device measurements, this methodology could be adapted to detect periodic structure in other scientific domains, including neuroscience, finance, and meteorology.

\vspace{-0.2em}






\section*{Supplementary Materials}

Supplementary materials available online include additional simulation details, convergence diagnostics, and posterior predictive checks. The \textit{Julia} files used for the analysis can be accessed at
\url{https://github.com/Beniamino92/BayesFreqSelect}.
Real data used in the applications can be obtained from the corresponding author, upon reasonable request.

	





\newpage 


\clearpage
\appendix

\renewcommand{\thesection}{S\arabic{section}}
\renewcommand{\thesubsection}{\thesection.\arabic{subsection}}
\renewcommand{\thefigure}{S\arabic{figure}}
\renewcommand{\thetable}{S\arabic{table}}
\renewcommand{\theequation}{S\arabic{equation}}

\setcounter{section}{0}
\setcounter{subsection}{0}
\setcounter{figure}{0}
\setcounter{table}{0}
\setcounter{equation}{0}

\begin{center}
    {\LARGE\bfseries Supplementary Material}
\end{center}

\vspace{1em}

\begin{abstract}
This document includes supplemental material to the article
``Frequency Selection in Bayesian Spectral Modeling of Time Series Data
with Applications to Wearable Device Measurements.'' Section~\ref{sec:suppl_MCMC_multi} describes the MCMC algorithm used to
extend the univariate spike-and-slab sampler to the multivariate setting.
Section~\ref{sec:s_convergence} presents convergence diagnostics for the
MCMC sampler, including trace plots and the Heidelberger and Welch
diagnostic test, and examines an alternative initialization strategy and
its impact on computational efficiency.
Section~\ref{sec:resolution_study_cfrequencies} investigates the ability
of the proposed method to distinguish closely spaced oscillatory
frequencies under different sample sizes and minimum-separation
constraints.
Section~\ref{sec:misspecified_simulation} evaluates the method under
model misspecification, considering data generated from an
autoregressive process and data with heavy-tailed, t-distributed
innovations.
Section~\ref{sec:sensitivity_analysis} examines the sensitivity of the
frequency-selection results to the hyperparameter governing the prior
inclusion probabilities.
Section~\ref{sec:s_mispecified} presents posterior predictive checks for
the misspecified autoregressive and heavy-tailed simulation settings.
Section~\ref{sec:suppl_bivariate_corr} examines a bivariate setting with
correlated errors.
Section~\ref{sec:s_wearable_all} presents the complete
frequency-selection results for all six individuals in the univariate
wrist actigraphy application.
Finally, Section~\ref{sec:s_wereable} presents posterior predictive
checks for both the univariate and bivariate real-data applications,
demonstrating the model's ability to closely reproduce the observed time
series.
\end{abstract}

	\section{MCMC Sampler for the Multivariate Model} \label{sec:suppl_MCMC_multi}

	In this section, we describe the MCMC algorithm used to extend the univariate spike-and-slab sampler to the multivariate setting, as introduced in Section~3.1 of the main paper. Each of the \( D \) time series is modeled as a sum of frequency-specific sine and cosine components, with its own set of coefficients and residual variance. A shared hierarchical prior links the frequency inclusion patterns across series, allowing the model to borrow strength while retaining component-specific flexibility.

	Let \( \boldsymbol{\beta} = (\boldsymbol{\beta}_1^\top, \dots, \boldsymbol{\beta}_{M_{\max}}^\top)^\top \) denote the full vector of Fourier coefficients across all components and candidate frequencies, where each \( \boldsymbol{\beta}_j = (\beta_{j1}^{(1)}, \beta_{j2}^{(1)}, \dots, \beta_{j1}^{(D)}, \beta_{j2}^{(D)})^\top \in \mathbb{R}^{2D} \) collects the sine and cosine weights of frequency \( \omega_j \in \Omega = \{ \tilde{\omega}_1, \dots, \tilde{\omega}_{M_{\max}} \} \) for all \( D \) components. Let \( \mathbf{z}_j = (z_{j1}, \dots, z_{jD}) \in \{0,1\}^D \) denote the vector of inclusion indicators for frequency \( \omega_j \) across components, and let \( \boldsymbol{\pi} \) be the prior distribution over these joint patterns. Define \( \mathbf{Z} = (\mathbf{z}_1, \dots, \mathbf{z}_{M_{\max}}) \) to be the collection of all inclusion indicators and let \( \boldsymbol{\sigma}^2 = (\sigma_1^2, \dots, \sigma_D^2) \) denote the component-specific noise variances. Then, Bayesian inference is based on the joint posterior distribution:
	\begin{equation}
		p(\mathbf{Z}, \boldsymbol{\beta}, \boldsymbol{\sigma}^2, \boldsymbol{\pi} \mid \mathbf{Y}, \boldsymbol{\omega}) \propto p(\mathbf{Y} \mid \mathbf{Z}, \boldsymbol{\beta}, \boldsymbol{\sigma}^2, \boldsymbol{\omega}) \cdot p(\mathbf{Z} \mid \boldsymbol{\pi}) \cdot p(\boldsymbol{\beta} \mid \mathbf{Z}) \cdot p(\boldsymbol{\sigma}^2) \cdot p(\boldsymbol{\pi}),
		\label{eq:multivariate_joint_posterior}
	\end{equation}
	where \( \mathbf{Y} = (\mathbf{y}^{(1)}, \dots, \mathbf{y}^{(D)}) \in \mathbb{R}^{T \times D} \) denotes the observed multivariate time series. The likelihood \( p(\mathbf{Y} \mid \cdot) \) factorizes over dimensions given the inclusion indicators and coefficients, and each coefficient vector \( \boldsymbol{\beta}_j \) is active only in the dimensions for which \( z_{jd} = 1 \). Let \( \boldsymbol{\beta}_{\mathbf{Z}} = \{ \boldsymbol{\beta}_j^{(d)} : z_{jd} = 1, \, j=1,\dots,M_{\max},\, d=1,\dots,D \} \) collect the active coefficients, and let \( \boldsymbol{\omega}_{\mathbf{Z}}^{(d)} = \{ \tilde{\omega}_j : z_{jd} = 1 \} \) denote the set of active frequencies for component \( d \).
	
	\medskip
	
	We devise a Metropolis-within-Gibbs sampler that iteratively alternates between sampling the inclusion patterns \( \mathbf{Z} \), the Fourier coefficients \( \boldsymbol{\beta}_{\mathbf{Z}} \), the residual variances \( \boldsymbol{\sigma}^2 \), and the hierarchical prior vector \( \boldsymbol{\pi} \), with proposals restricted to configurations that satisfy the minimum-separation constraint for each component. At each iteration, a joint inclusion pattern \( \mathbf{z}_j \) is updated via add, delete, or swap moves in a randomly selected component, followed by updates to the affected coefficients, variances, and the pattern prior. The full MCMC algorithm is detailed below.

	\paragraph*{$\bullet$ Jointly updating active frequency indicators  and linear basis coefficients.} 
	A proposal \( \mathbf{z}_j^{\text{prop}} \) is generated by modifying a single entry of \( \mathbf{z}_j \), corresponding to an add, delete, or swap move in one of the components. The move is accepted only if the proposed inclusion pattern respects the minimum-separation constraint in each affected component:
	\[
	\min_{\substack{k < \ell \\ z_{kd} = z_{\ell d} = 1}} |\omega_k - \omega_\ell| \geq \delta \quad \text{for all } d \in \{1, \dots, D\}.
	\]
	Let \( \mathcal{A} \subset \{1, \dots, D\} \) denote the set of components affected by the move. The acceptance probability is given by:
	\[
	\min\left(1, \exp\left\{ \log \frac{p(\mathbf{z}_j^{\text{prop}} \mid \boldsymbol{\pi})}{p(\mathbf{z}_j^{\text{curr}} \mid \boldsymbol{\pi})} + \sum_{d \in \mathcal{A}} \left[ \log p(\mathbf{y}^{(d)} \mid \mathbf{z}_j^{\text{prop}}, \boldsymbol{\beta}^{(d)}, \sigma_d^2) - \log p(\mathbf{y}^{(d)} \mid \mathbf{z}_j^{\text{curr}}, \boldsymbol{\beta}^{(d)}, \sigma_d^2) \right] \right\} \right),
	\]
	where the likelihood ratio is evaluated only for the affected dimensions, and the proposed coefficients \( \boldsymbol{\beta}^{(d)} \) are sampled from the conjugate Gaussian posterior conditional on the proposed set of active frequencies.
	
	\paragraph*{$\bullet$ Updating the active linear basis coefficients.}
	For each component \( d \), let \( M_d = \sum_{j=1}^{M_{\max}} z_{jd} \) denote the number of active frequencies. The corresponding Fourier coefficients are updated from the conjugate Gaussian posterior:
	\[
	\boldsymbol{\beta}_{\mathbf{z}}^{(d)} \sim \mathcal{N}_{2M_d}(\boldsymbol{\lambda}_T^{(d)}, \boldsymbol{\Sigma}_T^{(d)}),
	\]
	where
	\[
	\boldsymbol{\Sigma}_T^{(d)\,-1} = \frac{1}{\sigma_\beta^2} \mathbf{I}_{2M_d} + \frac{1}{\sigma_d^2} \mathbf{X}_d^\top \mathbf{X}_d, \qquad
	\boldsymbol{\lambda}_T^{(d)} = \frac{1}{\sigma_d^2} \boldsymbol{\Sigma}_T^{(d)} \mathbf{X}_d^\top \mathbf{y}^{(d)}.
	\]
	Here, \( \mathbf{X}_d \in \mathbb{R}^{T \times 2M_d} \) is the matrix of Fourier basis functions corresponding to the active frequencies in component \( d \), constructed as in the univariate case.
	
	\paragraph*{$\bullet$ Updating the residual variance.}
	Each residual variance \( \sigma_d^2 \) is updated independently from its inverse-gamma posterior:
	\[
	\sigma_d^2 \sim \text{IG}\left( \frac{T + \gamma_0}{2}, \frac{\nu_0 + \sum_{t=1}^T \left(y_t^{(d)} - \mathbf{x}_t^{(d)\top} \boldsymbol{\beta}_{\mathbf{z}}^{(d)} \right)^2}{2} \right).
	\]
	
	\paragraph*{$\bullet$ Updating the Pattern Inclusion Prior.}
	The prior vector \( \boldsymbol{\pi} \sim \text{Dirichlet}(\boldsymbol{\alpha}) \) governs the distribution over all possible inclusion configurations \( \mathbf{z}_j \in \{0,1\}^D \). At each iteration, we update \( \boldsymbol{\pi} \) from its conjugate posterior:
	\[
	\boldsymbol{\pi} \mid \{\mathbf{z}_j\} \sim \text{Dirichlet}\left(\alpha_1 + C_1, \dots, \alpha_{2^D} + C_{2^D} \right),
	\]
	where \( C_h \) denotes the number of frequencies currently assigned to configuration \( h \), corresponding to the \( h \)-th binary vector in \( \{0,1\}^D \).
	
	\vspace{0.3cm}
	
	The multivariate sampler introduces several key extensions relative to its univariate counterpart. While the univariate model assigns a scalar inclusion indicator $z_j$
	to each frequency, the multivariate model uses a vector $\bm{z}_j$, allowing for frequency-specific inclusion across multiple components. These joint inclusion patterns are governed by a shared prior 
	$\bm{\pi}$ which encourages coordinated frequency selection across components when supported by the data. In addition, the multivariate model explicitly updates 
	$\bm{\pi}$ at each iteration via its conjugate Dirichlet posterior, a step that is absent in the univariate setting. The proposal mechanism is also adapted: add and delete moves modify the inclusion indicator of a single component at a time, whereas swap moves may involve exchanging inclusion status across different components, offering richer exploration of the joint model space. In all cases, proposals are accepted only if they satisfy the minimum-separation constraint within each component. Furthermore, unlike the univariate model which includes a single residual variance, the multivariate formulation models a separate variance $\sigma_d$
	for each series, allowing component-specific noise modeling. Finally, to improve computational efficiency, the likelihood ratio in the Metropolis–Hastings step is evaluated only over the subset of components affected by the proposed change, in contrast to the univariate case where the full likelihood is always recomputed.

	\section{Simulation Study. Convergence Diagnostic} \label{sec:s_convergence}
	
\begin{figure}[ht!]
		\centering
		\includegraphics[width=0.6\textwidth]{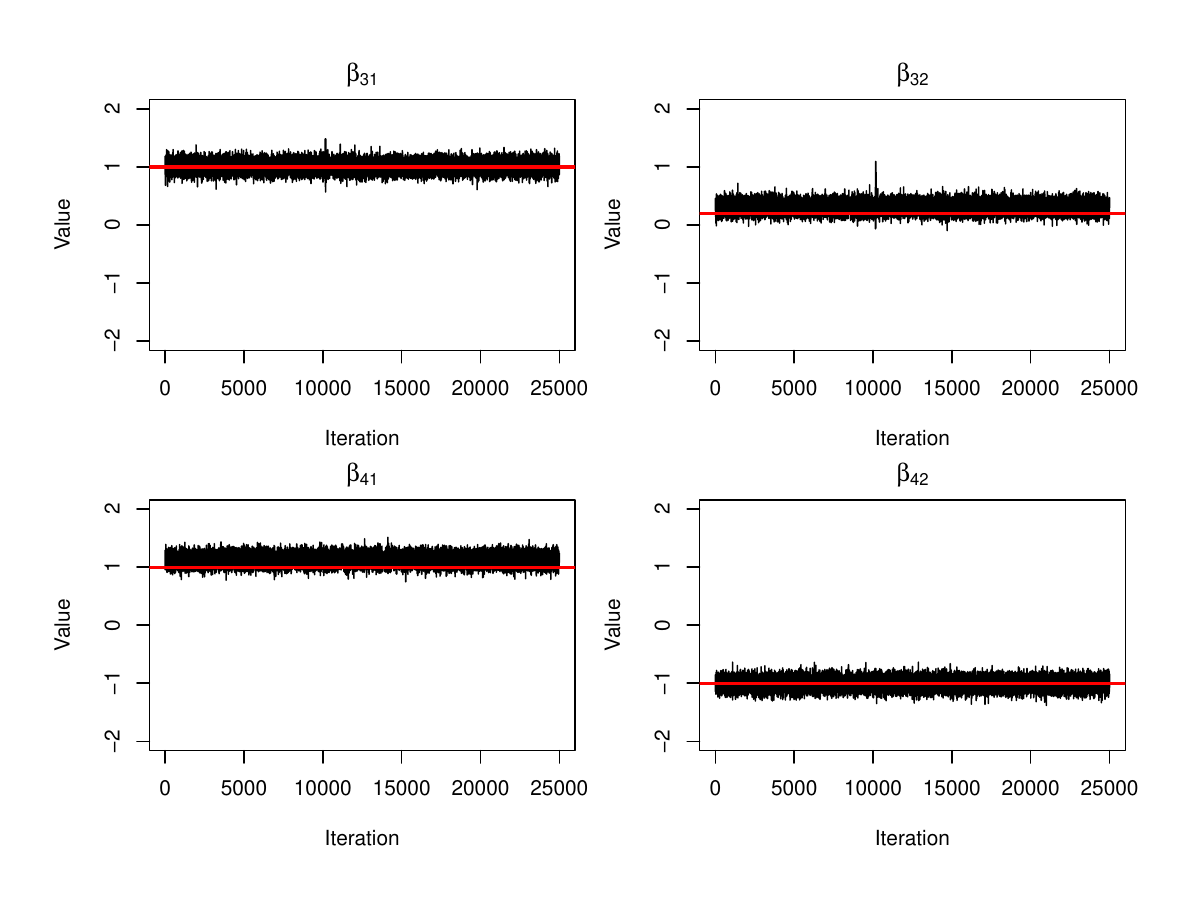}
		\caption{\textbf{Simulation Study.} Representative examples of trace plots for the linear coefficients $\beta_{31}, \beta_{32}, \beta_{41},$ and $\beta_{42}$}
		\label{fig:traces_beta}
		\vspace{1cm} 
		
		\includegraphics[width=0.6\textwidth]{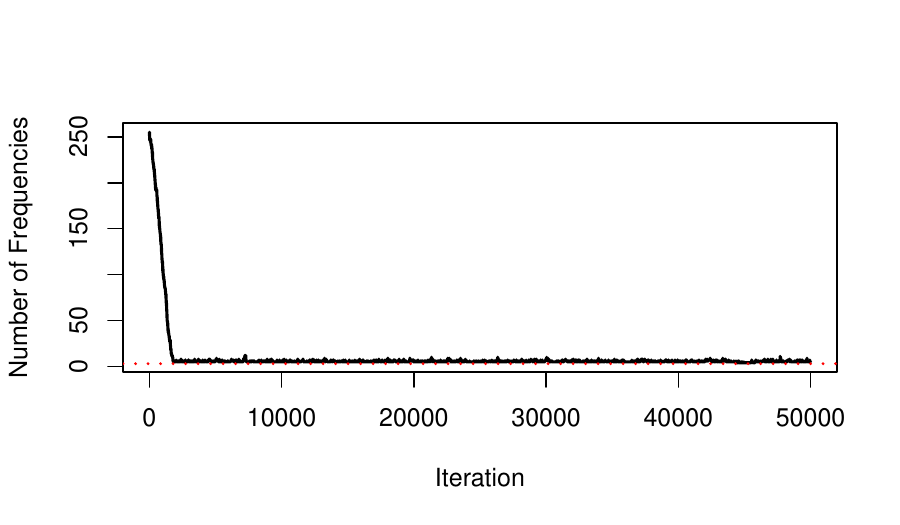}
		\caption{\textbf{Simulation Study.} Number of frequencies selected in the model across MCMC iterations, with the red horizontal dotted line indicating the true number of generating frequencies }
		\label{fig:traces_nFreq}
	\end{figure}

	In our experiments, to ensure the MCMC sampler had converged, we performed the following checks: (i) we examined trace plots of the parameters and observed no evidence of pathological behavior (representative examples of trace plots for the linear coefficients $\beta_{31}, \beta_{32}, \beta_{41},$ and $\beta_{42}$ are shown in Figure \ref{fig:traces_beta}); (ii) we applied the Heidelberger and Welch convergence diagnostic test \citep{heidelberger1981spectral}. This diagnostic evaluates whether the Markov chain has reached its stationary distribution and subsequently assesses if the mean of the marginal posterior distribution can be estimated with adequate precision under the assumption of stationarity. In our experiments, all analyzed Markov chains successfully passed this diagnostic.
	
	In the simulation study described in Section 4.1, we examined an alternative initialization strategy that included all frequencies in the model from the outset, using the set of Fourier frequencies for illustration rather than a refined grid. This approach was designed to evaluate the sampler’s ability to accurately identify the relevant frequencies and their associated linear coefficients. Figure \ref{fig:traces_nFreq} illustrates the number of frequencies selected in the model across MCMC iterations, with the red horizontal dotted line indicating the true number of generating frequencies. The results show that the model stabilizes around the correct number of frequencies after approximately 2000 iterations. While this method proved effective, initializing the sampler with just the two peaks significantly reduced the computational time needed for convergence.

	\section{Simulation Study: Resolution Study for Closely Spaced Frequencies} \label{sec:resolution_study_cfrequencies}
	
    We further investigate the behavior of the proposed method in scenarios where oscillatory components are closely spaced in the frequency domain, and compare its performance with the alternative frequency selection methods considered earlier. In particular, we consider settings where the separation between true frequencies may be smaller than the minimum spacing imposed by the constraint introduced in the main paper. Recall that this constraint requires any two active frequencies to satisfy $|\omega_j - \omega_k| \ge d \Delta \omega$.  
	
	We generated time series of length $T \in \{128, 256, 512\}$ from a model with two frequencies components whose separation decreases with the sample size. Specifically, the two true frequencies were defined as $\omega_1 = 0.10$ and $\omega_2 = \omega_1 + \delta$, where the separation parameter is given by $\delta = c/T$. We considered values $c \in \{1, 2, 5\}$, producing increasingly small separations between the two oscillatory components as $T$ grows. This design creates increasingly challenging regimes in which the two frequencies become progressively harder to distinguish. The corresponding regression coefficients were set to $(\beta_{11},\beta_{12}) = (0.6,-0.6)$ and $(\beta_{21},\beta_{22}) = (1.0,0.3)$, while the residual variance was fixed at $\sigma^2 = 1$. For each configuration we generated 20 independent datasets. For the proposed spectralSS method, we explored several values of the minimum separation parameter $d \in \{0,1,2,5\}$ to assess the impact of this constraint. The candidate frequency grid was defined over $(0,0.5)$ with spacing $\Delta \omega = 0.0001$, matching the setup used in the previous simulation study.
	
\begin{figure}
		\centering
			\includegraphics[width=0.9\textwidth]{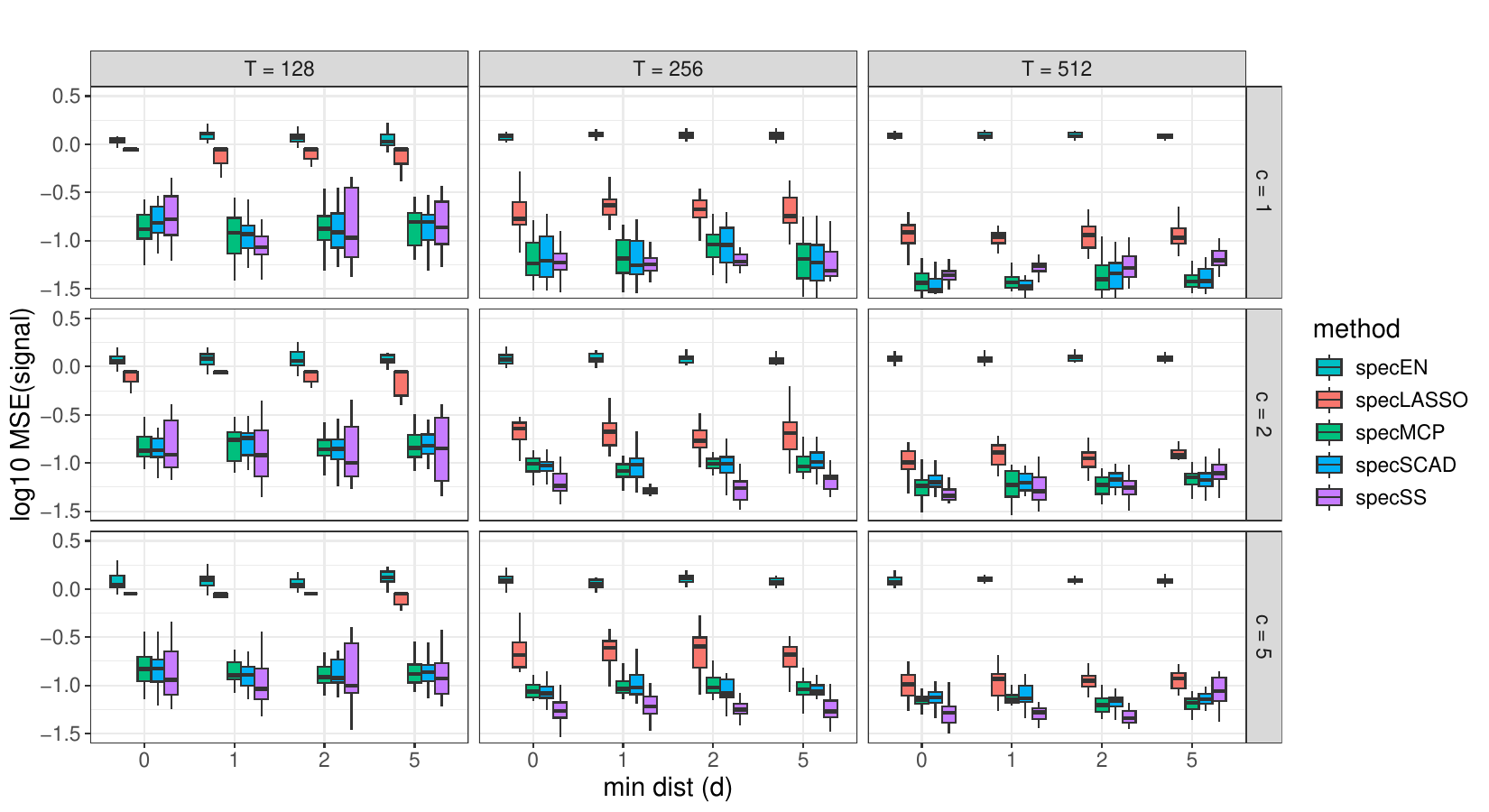}
			\includegraphics[width=0.9\textwidth]{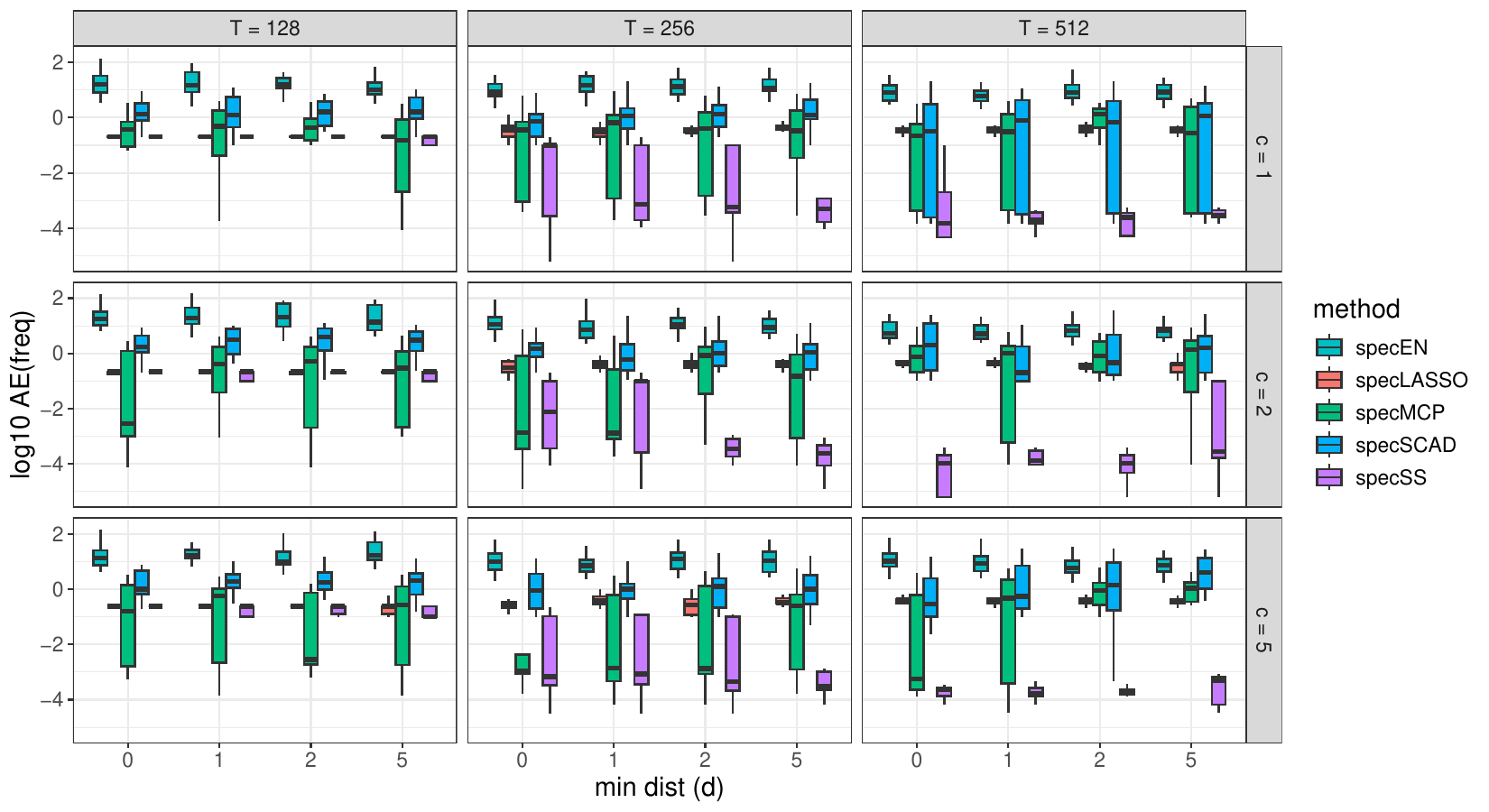}
		\caption{\textbf{Simulation Study.} 
                Performance comparison of frequency selection methods under closely spaced frequencies. (Top) Boxplots of $\log_{10}(MSE_S)$. (Bottom) Boxplots of $\log_{10}(AE_F)$.   
				Results are shown for different sample sizes $T$, separation levels $\delta = c/T$, 
				and minimum spacing parameters $d$. Each boxplot summarizes results over 20 simulated datasets.}
		\label{fig:resolution}
	\end{figure}

	Performance was evaluated using the same metrics introduced in the main paper. In particular, we report $MSE_S$ and $AE_F$.
	The results are summarized in Figure \ref{fig:resolution}. The top panel reports the signal reconstruction error $MSE_S$. Across most configurations of $T$, $c$, and the separation parameter $d$, the proposed spectralSS method achieves smallest $MSE_S$ values. The differences between methods become less pronounced as the sample size increases. For example, when $T=512$ and $d=5$, the specMCP and specSCAD approaches occasionally achieve slightly smaller reconstruction errors. However, in the majority of scenarios spectralSS yields the lowest $MSE_S$ across repeated datasets. As expected, the variability of the estimates is larger for smaller sample sizes. The bottom panel of Figure \ref{fig:resolution} reports the frequency estimation error $AE_F$. Under this metric, spectralSS consistently produces the smallest errors across most combinations of $T$, $c$, and $d$. The specMCP method sometimes achieves comparable performance, but its results exhibit substantially larger variability across simulated datasets. Overall, these results indicate that the proposed spectralSS approach achieves smaller $MSE_S$ and $AE_F$ values than the competing approaches, even in challenging scenarios that feature regimes with small separation. These findings provide empirical insight into the resolution limits of the proposed
	method and the minimum data requirements needed to reliably distinguish nearby frequencies.
	
\section{Simulation Study: Misspecified Model} \label{sec:misspecified_simulation}

We examine the effectiveness of our proposed method in detecting spectral peaks when there is a mismatch between the model and the data-generating process. Specifically, we conducted simulation studies under two distinct conditions. In the first scenario, data were generated from an autoregressive (AR) process, allowing us to compare our method’s performance with that of SpectralLASSO and SpectralEN. In the second scenario, we generated data with t-distributed innovations, hence violating the Gaussian assumption for $\varepsilon_t$ in the assumed generative model. For both settings, we select a grid of Fourier frequencies, set \( b = 100 \) for the spectralSS beta parameter, and use the remaining parameters as specified in the main paper. Posterior predictive graphical checks are included in the Supplementary Material.

\subsection{Autoregressive Process}
In the signal processing literature, it is common to model a time series as a linear combination of a finite set of sinusoids with added noise; however, such line-spectrum models are relatively uncommon in the statistics literature. Instead, it is generally assumed that the power spectrum varies smoothly across frequencies. Here, we assess the performance of our proposed procedure on data generated from an AR process with a spectral density function exhibiting a sharp peak. Specifically, we simulated 50 realizations from the following AR process
\begin{equation}
	y_t = 1.9y_t-1 - 0.975 y_{t-2} + \varepsilon_t, 
	\label{eq:AR}
\end{equation}
where $\varepsilon_t \sim \mathcal{N}(0, 0.25)$. Figure \ref{fig:mispecified} (a) displays a realization for model \eqref{eq:AR}. The average number of estimated frequencies was 80.5, 2.33, and 70.9 for SpectralEN, spectralSS, and SpectralLASSO, respectively, indicating that our proposed method is considerably more parsimonious in identifying the primary peak driving the overall variation in the data. Figure \ref{fig:mispecified} (c) presents boxplots of $\log(AE_F)$ for each method. It is evident that, even in this scenario, where the data-generating model differs significantly from our model’s underlying assumptions, our approach appears to outperform spectralEN and spectralLASSO in estimating the frequency peak.

\subsection{Non-Gaussian Error}
\begin{figure}
	\centering
		\includegraphics[width=2.5in]{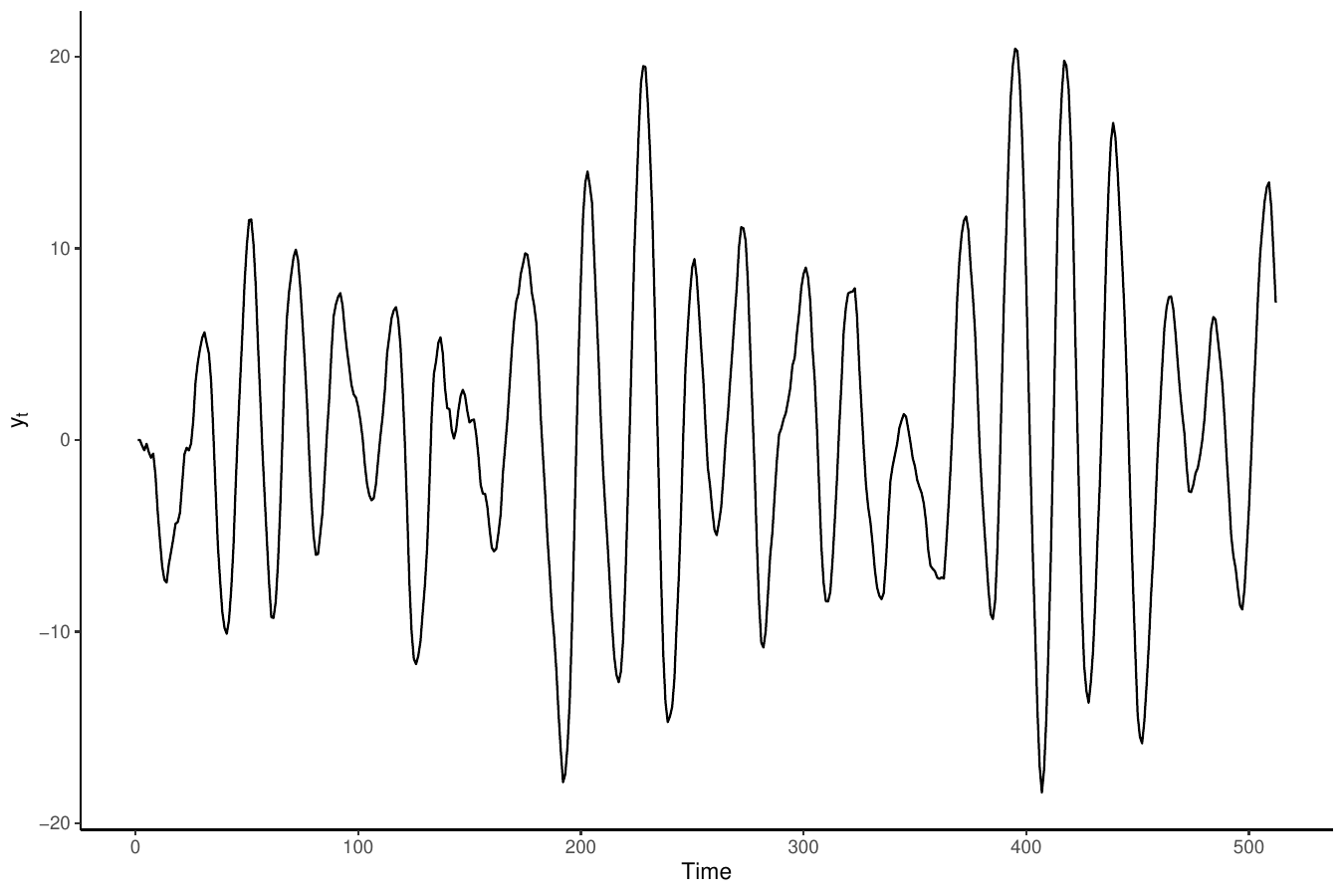}
		\includegraphics[width=2.5in]{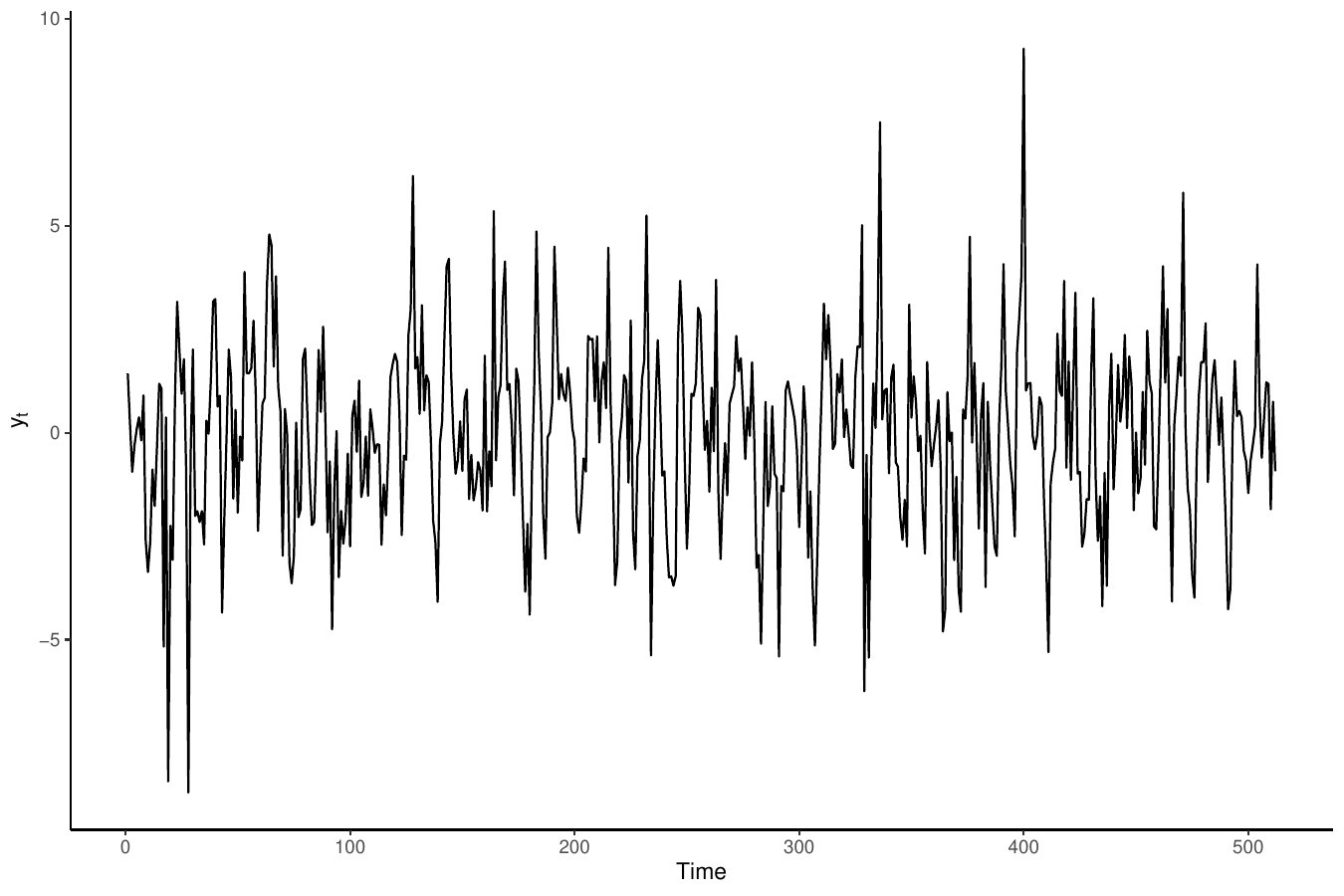}
		\includegraphics[width=2.5in]{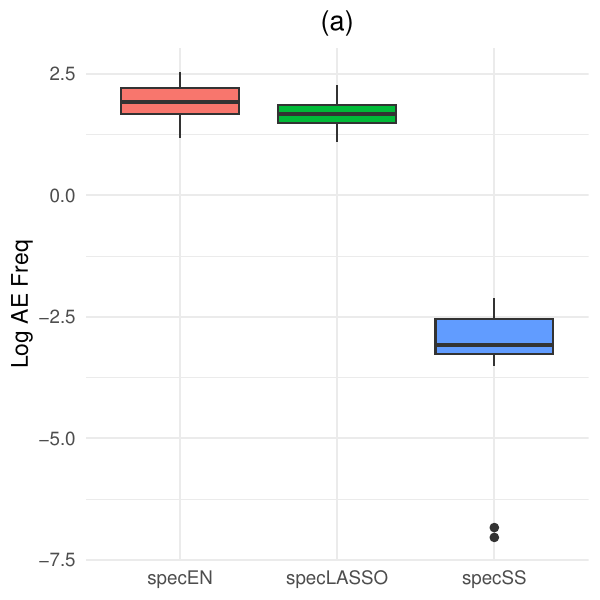}
		\includegraphics[width=2.5in]{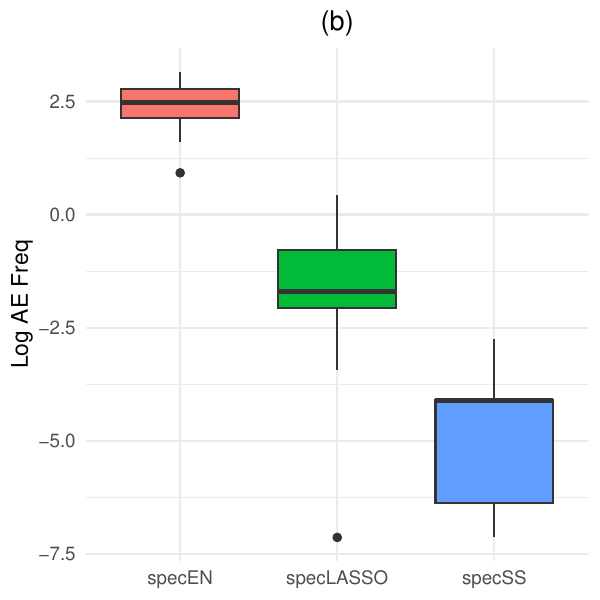}
		\caption{}
		\label{fig:subfig4}
	\caption{\textbf{Simulation Study}.  Sample realizations from (a an AR process and (b) a process with t-distributed innovations.  Boxplots of $\log(AE_F)$ for spectralSS, spectralEN, and spectralLASSO under (c) AR process data and (d) t-distributed innovations.}
	\label{fig:mispecified}
	
\end{figure}

We evaluate our approach’s performance in settings where the innovations follow a t-distribution. Specifically, we simulate a time series using the same simulation setting described in Section 4.1, this time generating errors from a t-distribution with 3 degrees of freedom. This choice of degrees of freedom introduces heavy-tailed behavior in the distribution, allowing us to assess the method’s robustness under conditions of higher variability. An example realization of this time series is illustrated in Figure \ref{fig:mispecified} (b). The average estimated frequencies were 61.13 for SpectralEN, 3.26 for spectralSS, and 6.13 for SpectralLASSO. These results suggest that our proposed method is substantially more parsimonious in identifying the correct number of peaks responsible for the main variation in the data, whereas SpectralEN tends to considerably overestimate the number of peaks. Figure \ref{fig:mispecified} (d) shows boxplots of $\log(AE_F)$ for each method, illustrating that spectralSS seems to perform better than the other two approaches in this example. 
In conclusion, although our model is based on the assumption of Gaussianity, spectralSS appears to perform effectively even when the underlying oscillatory process follows a heavy-tailed t-distribution.


\section{Simulation Study: Sensitivity Analysis} \label{sec:sensitivity_analysis}
We carried out a sensitivity analysis study focusing on the impact of the hyperparameter $b$ of the inclusion probability $ \alpha_j \sim \text{Beta}(a, b) $, recalling that the latent inclusion indicator variable for frequency $ \omega_j $ follows a Bernoulli distribution with success parameter $ \alpha_j $. We simulated 20 time series from the same simulation setting of Section 4.1, for different values of $b \in \{1, 5, 10, 100, 1000, 10000, 100000 \}$ and different sample size $T \in \{250, 500, 1000, 5000 \}$. Table \ref{tab:sensitivity_analysis} displays $AE_F$, $AE_P$, $MSE_S$ and $\hat{M}$, for each scenarios.
As expected, the number of estimated frequencies, $\hat{M}$, decreases as $b$ increases. This is due to the stronger regularization effect introduced by a higher $b$ value, which leads to a greater probability of non-inclusion for frequencies in the model. Consequently, as $b$ becomes larger, fewer frequencies are included, resulting in a lower $\hat{M}$. This reflects a higher degree of shrinkage, where only the most significant frequencies are retained in the model.
Furthermore, the impact of the prior is more pronounced for smaller sample sizes. For $T = 250$ and $T = 500$, we observe greater variability in the metrics $AE_F$, $AE_P$, and $MSE_S$. This suggests that, with smaller sample sizes, the prior plays a more critical role in shaping the posterior distribution, which influences the model's estimation accuracy and prediction performance. In contrast, as the sample size increases to $T = 1000$ and $T = 5000$, the model’s performance becomes more stable, and the effect of the prior diminishes. With larger datasets, the data itself has more influence over the estimation of the inclusion probabilities and frequencies, reducing the sensitivity to the hyperparameter $b$.

\begin{table}[ht!]
	\caption{\textbf{Simulation Study.} Sensitivity analysis showing the effect of the inclusion probability hyperparameter $b$ on $AE_F$, $AE_P$, $MSE_S$, and $\hat{M}$ for different values of $b \in \{1, 5, 10, 100, 1000, 10000, 100000\}$ and sample sizes $T \in \{250, 500, 1000, 5000\}$. }
	\centering
	\resizebox{\textwidth}{!}{%
		\begin{tabular}{lcccclllcccc} \hspace{-0.2cm}\\ \hline
			& \multicolumn{4}{c}{$T = 250$}                                                               &  &  &            & \multicolumn{4}{c}{$T = 500$}      \\  \cmidrule{2-5} \cmidrule{9-12}                                                         \\
			& $AE_F$                 & $AE_P$                 & $MSE_S$                & $M$                    &  &  &            & $AE_F$                 & $AE_P$                 & $MSE_S$                & $M$                    \\
			$b = 1$      & 0.239                & 4.135                & 0.229                & 7.65                 &  &  & $b = 1$      & 0.042                & 1.337                & 0.151                & 7.75                 \\
			$b = 5$      & 0.158                & 3.143                & 0.222                & 3.90                 &  &  & $b = 5$      & 0.011                & 0.463                & 0.143                & 4.00                 \\
			$b = 10$     & 0.124                & 2.827                & 0.242                & 3.90                 &  &  & $b = 10$     & 0.003                & 0.458                & 0.163                & 4.00                 \\
			$b = 100$    & 0.150                & 3.092                & 0.383                & 3.45                 &  &  & $b = 100$    & 0.007                & 0.584                & 0.154                & 4.00                 \\
			$b = 1000$   & 0.189                & 3.604                & 0.861                & 2.00                 &  &  & $b = 1000$   & 0.003                & 0.462                & 0.162                & 4.00                 \\
			$b = 10000$  & 0.221                & 4.035                & 1.370                & 1.20                 &  &  & $b = 10000$  & 0.009                & 0.754                & 0.210                & 3.85                 \\
			$b = 100000$ & 0.246                & 4.189                & 1.482                & 1.00                 &  &  & $b = 100000$ & 0.007                & 0.594                & 0.264                & 3.70                 \\
			& \multicolumn{1}{l}{} & \multicolumn{1}{l}{} & \multicolumn{1}{l}{} & \multicolumn{1}{l}{} &  &  &            & \multicolumn{1}{l}{} & \multicolumn{1}{l}{} & \multicolumn{1}{l}{} & \multicolumn{1}{l}{} \\
			& \multicolumn{4}{c}{$T = 1000$}                                                              &  &  &            & \multicolumn{4}{c}{$T = 5000$}           \\ \cmidrule{2-5} \cmidrule{9-12}                                                    \\
			& $AE_F$                 & $AE_P$                 & $MSE_S$                & $M$                    &  &  &            & $AE_F$                 & $AE_P$                 & $MSE_S$                & $M$                    \\
			$b = 1$      & 0.013                & 0.239                & 0.148                & 8.30                 &  &  & $b = 1$      & 0.078                & 0.594                & 0.698                & 8.85                 \\
			$b = 5$      & 0.021                & 0.226                & 0.135                & 4.60                 &  &  & $b = 5$      & 0.071                & 0.518                & 0.625                & 5.40                 \\
			$b = 10$     & 0.021                & 0.186                & 0.141                & 4.45                 &  &  & $b = 10$     & 0.047                & 0.394                & 0.501                & 4.90                 \\
			$a = 100$    & 0.011                & 0.182                & 0.152                & 4.40                 &  &  & $b = 100$    & 0.025                & 0.437                & 0.487                & 4.40                 \\
			$a = 1000$   & 0.013                & 0.216                & 0.166                & 4.25                 &  &  & $b = 1000$   & 0.013                & 0.39                 & 0.435                & 4.20                 \\
			$a = 10000$  & 0.000                & 0.152                & 0.172                & 4.00                 &  &  & $b = 10000$  & 0.000                & 0.386                & 0.394                & 4.00                 \\
			$a = 100000$ & 0.000                & 0.211                & 0.17                 & 4.00                 &  &  & $b = 100000$ & 0.000                & 0.386                & 0.394                & 4.00      \\[0.1em] \hline               
	\end{tabular}}
	\label{tab:sensitivity_analysis}
\end{table}



	\section{Misspecified Model. Posterior Predictive} \label{sec:s_mispecified}
	
\begin{figure}[ht!]
		\centering
		\includegraphics[width=0.6\textwidth]{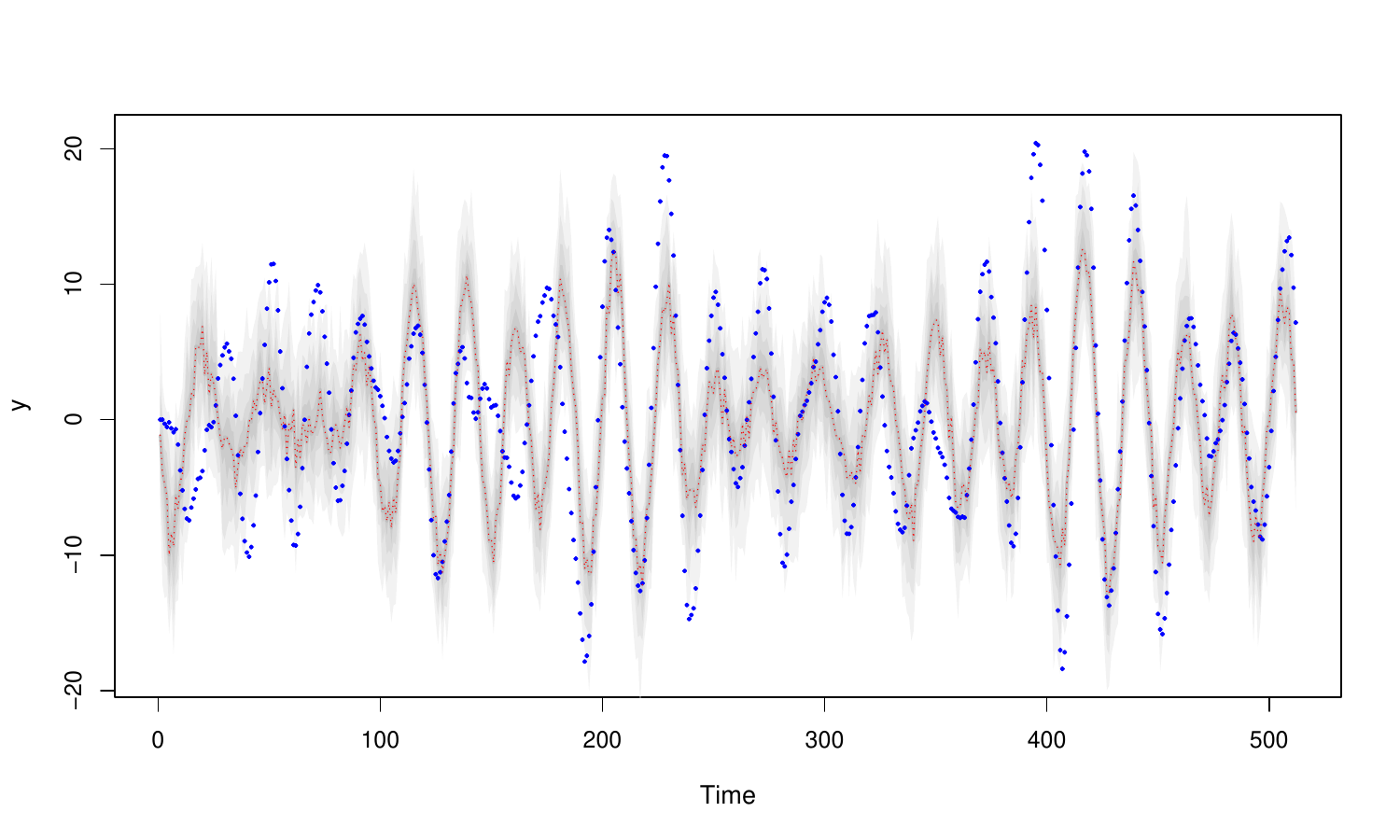}
		\caption{\textbf{Simulation Study.} Posterior predictive check for data generated from an AR process, consisting of a realization of the time series alongside the estimated signal (red line) and 30 posterior predictive draws. }
		\label{fig:postpred__AR}
		
		\vspace{1cm} 
		
		\includegraphics[width=0.6\textwidth]{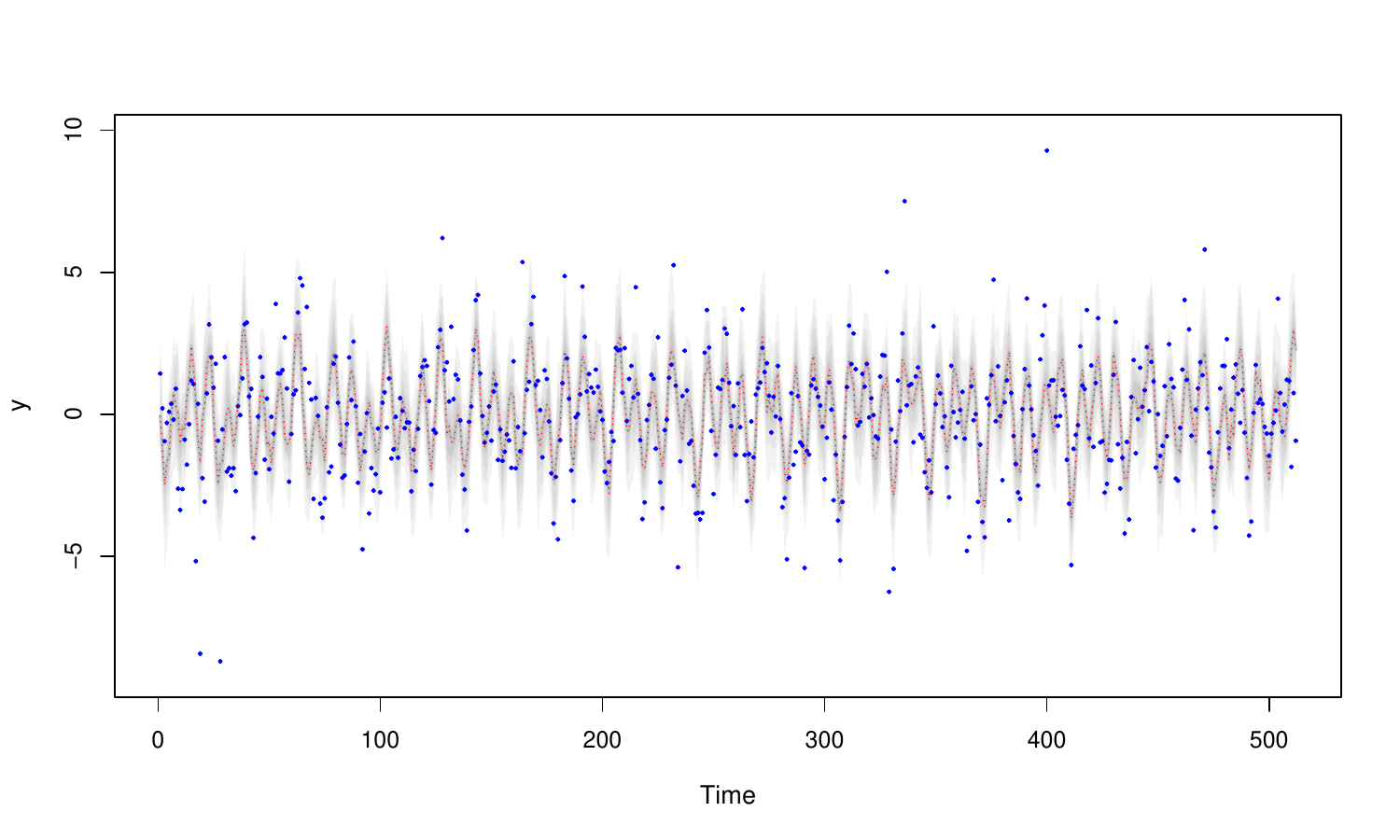}
		\caption{Posterior predictive check for data with t-distributed innovations, consisting of a realization of the time series alongside the estimated signal (red line) and 30 posterior predictive draws. }
		\label{fig:postpred_T}
	\end{figure}

	We present a posterior predictive check for two scenarios: data generated from an autoregressive (AR) process (Figure \ref{fig:postpred__AR}) and data with t-distributed innovations (Figure \ref{fig:postpred_T}), as described in Section 3.2. Each plot displays a realization of the time series alongside the estimated signal and 30 posterior predictive draws. Even when the model was misspecified, it produced satisfactory results, with the fit closely resembling the observed data.
	
\section{Bivariate Time Series with Correlated Errors} \label{sec:suppl_bivariate_corr}
		
\begin{figure}[ht!]
			\centering
			
			\captionsetup[subfigure]{labelformat=empty} 
			\makebox[\textwidth][c]{\textbf{(a)}} 
			
			\begin{minipage}[t]{0.48\textwidth}
				\centering
				\includegraphics[width=\linewidth]{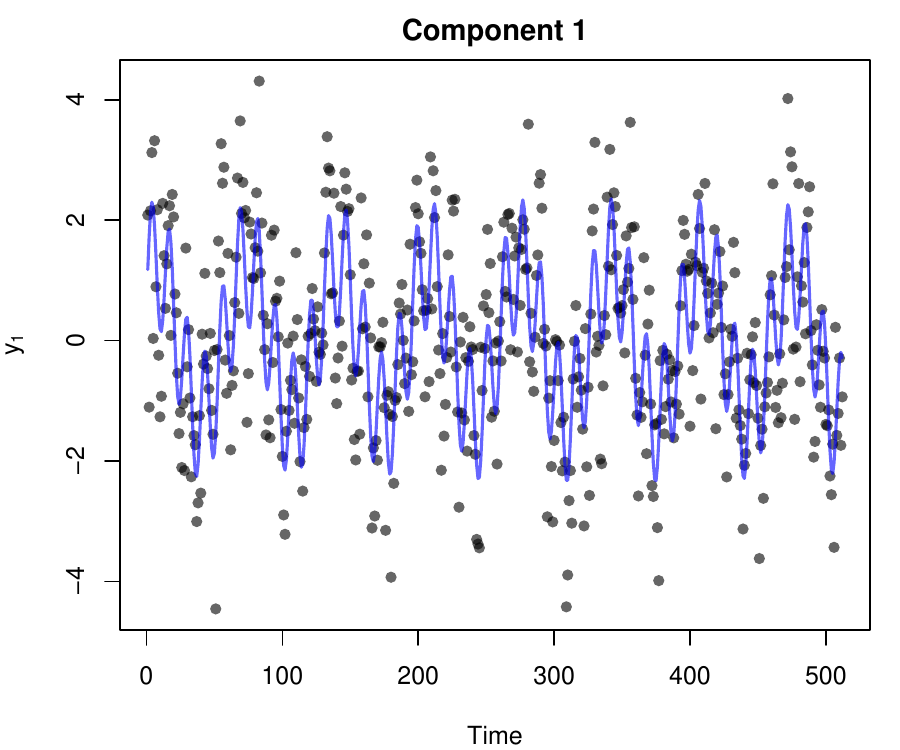}
			\end{minipage}
			\hfill
			\begin{minipage}[t]{0.48\textwidth}
				\centering
				\includegraphics[width=\linewidth]{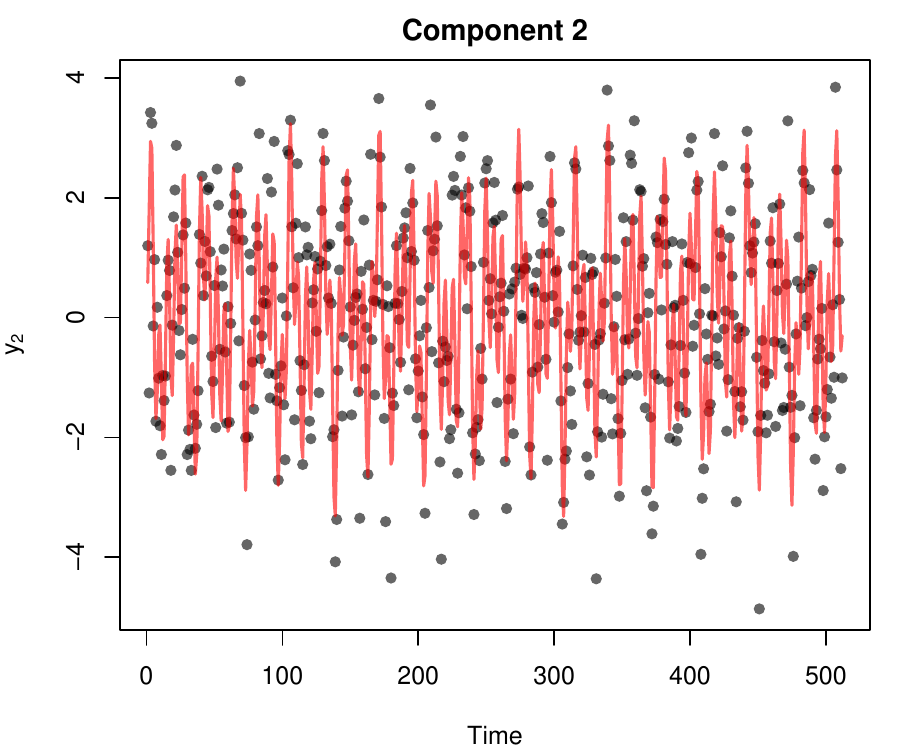}
			\end{minipage}

			\makebox[\textwidth][c]{\textbf{(b)}} 
			
			\begin{minipage}[t]{0.48\textwidth}
				\centering
				\includegraphics[width=\linewidth]{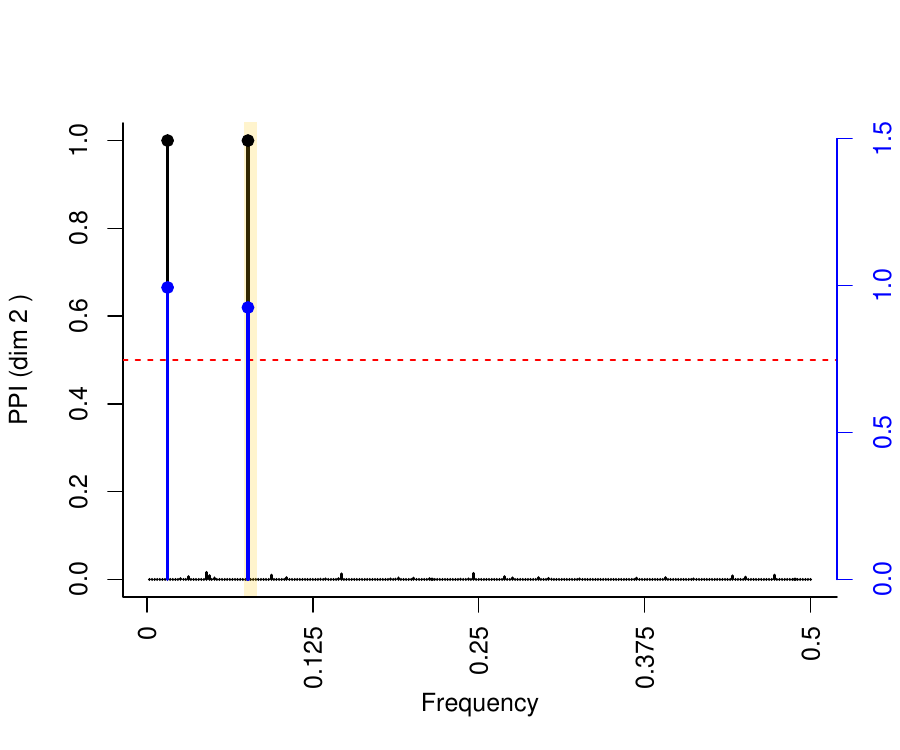}
			\end{minipage}
			\hfill
			\begin{minipage}[t]{0.48\textwidth}
				\centering
				\includegraphics[width=\linewidth]{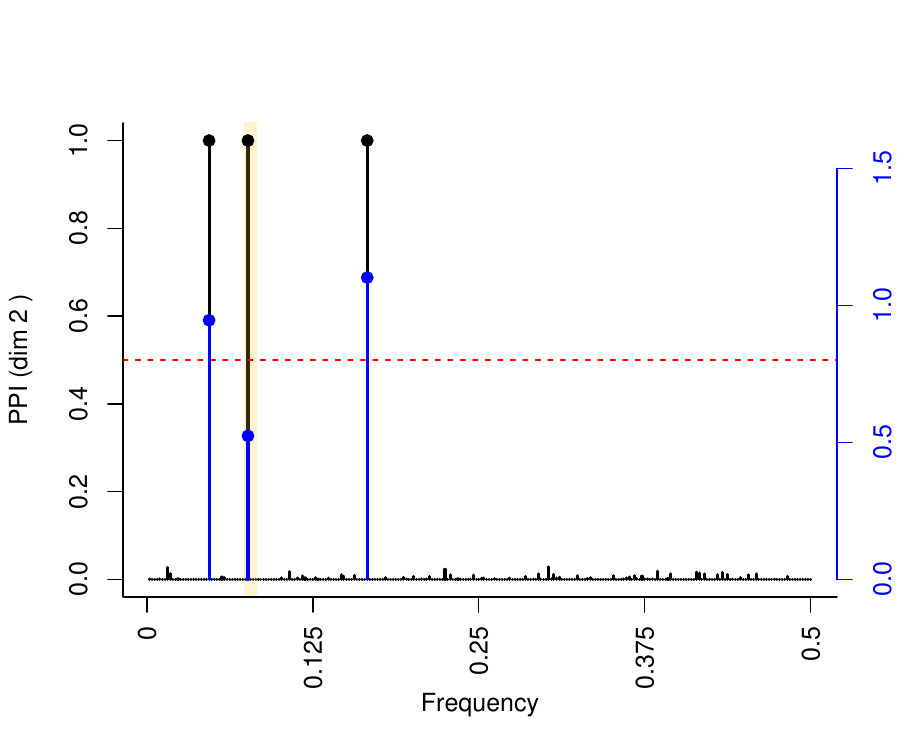}
			\end{minipage}
			
			\caption{\textbf{Bivariate Time Series with Correlated Errors } 
				(a) Simulated observations (dots) and true underlying signal (lines) for the two components. 
				(b) Posterior inclusion probabilities (PPIs; black lines) and estimated square root of signal power (blue lines) for frequencies with PPI greater than 0.5. 
				The shaded region highlights a frequency that is jointly selected across both components but with distinct amplitudes, illustrating the model’s ability to recover partially shared spectral features.}
			
			\label{fig:multiv_simul_correlated}
		\end{figure}

		We consider the same bivariate simulation setting as in Section 5.4, with \( T = 512 \) time points and identical frequency configurations, coefficients, and prior specifications. 
		The only difference is in the noise structure: instead of independent errors across components, we generate correlated Gaussian errors at each time point. Specifically, for each \( t \), we let
		\[
		(\varepsilon_{t1}, \varepsilon_{t2})^\top \sim \mathcal{N}(\mathbf{0}, \boldsymbol{\Sigma}), 
		\quad 
		\boldsymbol{\Sigma} =
		\begin{pmatrix}
			\sigma_1^2 & \rho\,\sigma_1\sigma_2 \\
			\rho\,\sigma_1\sigma_2 & \sigma_2^2
		\end{pmatrix},
		\]
		with \( \sigma_1^2 = \sigma_2^2 = 1 \) and \( \rho = 0.9 \), while the fitted model still assumes independent errors.
		Figure~\ref{fig:multiv_simul_correlated}(a) shows a realization of the process alongside the underlying signal.
		
		The posterior mode for the number of active frequencies remains \( M_1 = 2 \) and \( M_2 = 3 \), with posterior probabilities \( p(M_1 = 2 \mid \mathbf{y}, \cdot) = 0.8618 \) and \( p(M_2 = 3 \mid \mathbf{y}, \cdot) = 0.615 \), though with increased posterior mass on larger models, particularly for the second component. Conditional on these modal values, the estimated dominant frequencies are unchanged, with \( \boldsymbol{\omega}_1 = (0.0156, 0.0762) \) and \( \boldsymbol{\omega}_2 = (0.0469, 0.0762, 0.1660) \), closely matching the true generating values. The corresponding estimated signal power is \( P(\boldsymbol{\omega}_1) = (0.994, 0.925) \) and \( P(\boldsymbol{\omega}_2) = (0.946, 0.524, 1.10) \).
		
		Figure~\ref{fig:multiv_simul_correlated}(b) displays the PPIs across the candidate frequency grid for each component, along with the estimated square root of the signal power for frequencies with \(\mathrm{PPI} > 0.5\). The model continues to accurately recover the true frequency structure, including the shared frequency at \( \omega \approx 0.076 \), despite the misspecified independence assumption, although with slightly increased uncertainty in model size selection.

\section{Applications to Wearable Device Data: Results for All Subjects}
\label{sec:s_wearable_all}

\begin{figure}[ht!]
	\centering
	\includegraphics[width=1.0\textwidth]{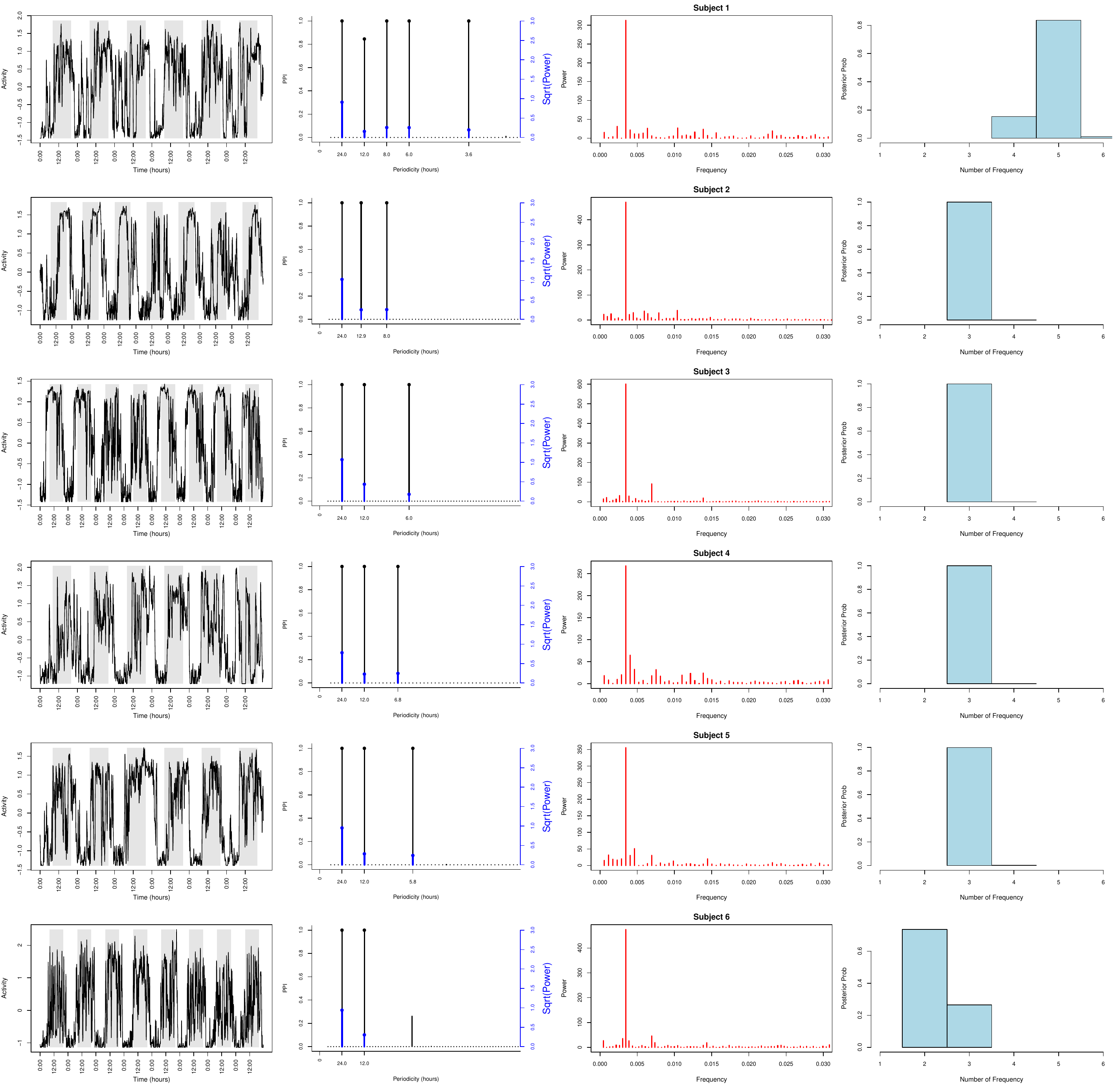}
\caption{\textbf{Wearable Activity Monitoring.} Results for all six individuals. The first column shows the standardized activity time series, with gray bands indicating 8 AM--8 PM; the second shows the estimated periodicities with posterior inclusion probabilities and square-root spectral power; the third shows the raw periodogram; and the fourth shows the posterior distribution of the number of selected frequencies.}
	\label{fig:results_data_all}
\end{figure}

To provide the complete results for the univariate wrist actigraphy application, Figure~\ref{fig:results_data_all} presents the analysis for all six individuals with partial-onset seizures. For each subject, the figure displays the standardized activity time series, the estimated periodicities with posterior inclusion probabilities and square-root spectral power, the raw periodogram, and the posterior distribution of the number of selected frequencies. The results reveal a dominant circadian periodicity across subjects, together with subject-specific ultradian components and varying levels of oscillatory complexity.
    
	\section{Applications to Wearable Device Data: Posterior Predictive} 
	\label{sec:s_wereable}

\begin{figure}[ht!]
		\centering
		\includegraphics[width=0.6\textwidth]{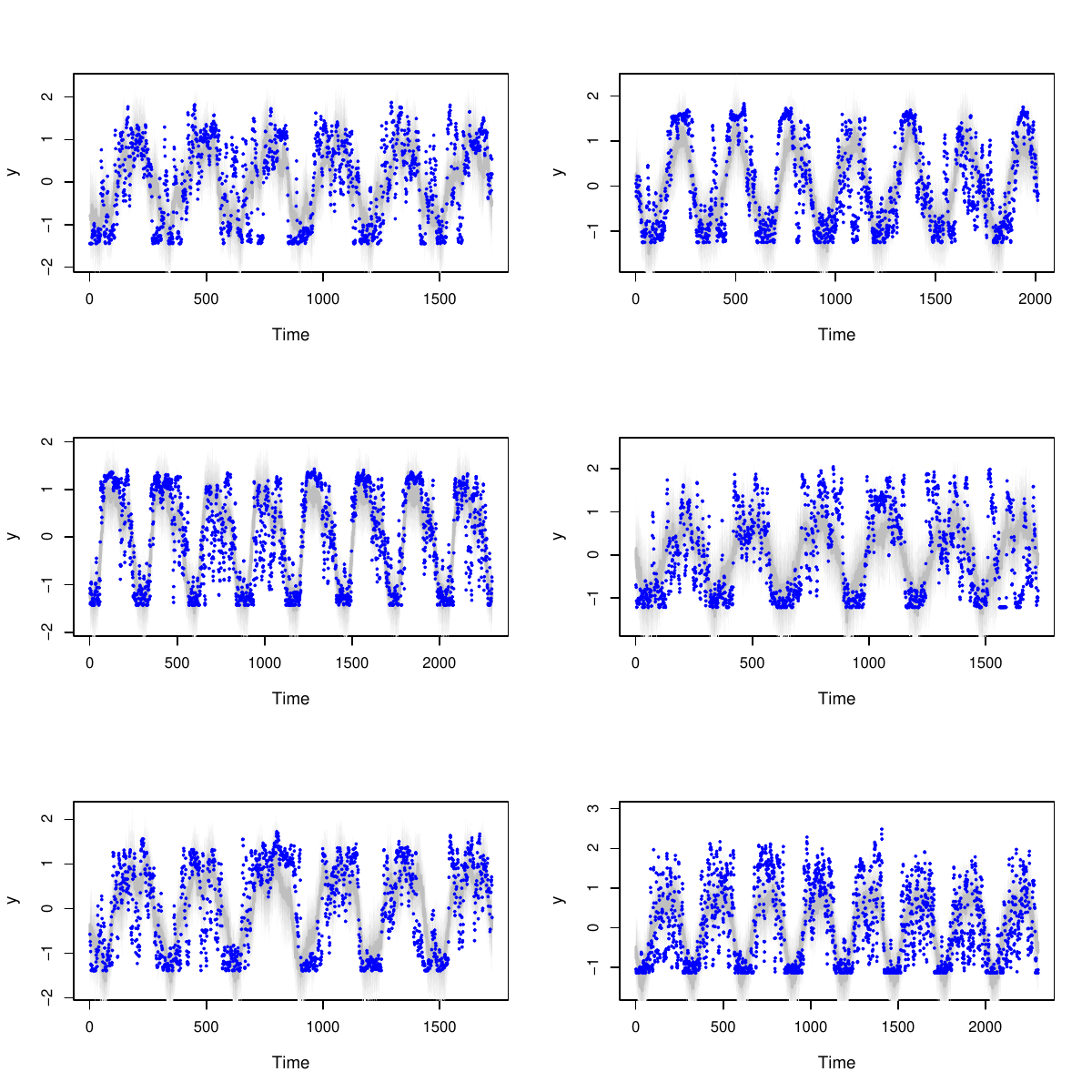}
		\caption{\textbf{Wearable Activity Monitoring (Univariate).} Posterior predictive checks for different patients with partial-onset seizures. Each plot shows the observed time series along with 30 posterior draws.}
		\label{fig:postpred_subjects_univar}
	\end{figure}

\begin{figure}[ht!]
		\centering
		\includegraphics[width=0.4\textwidth]{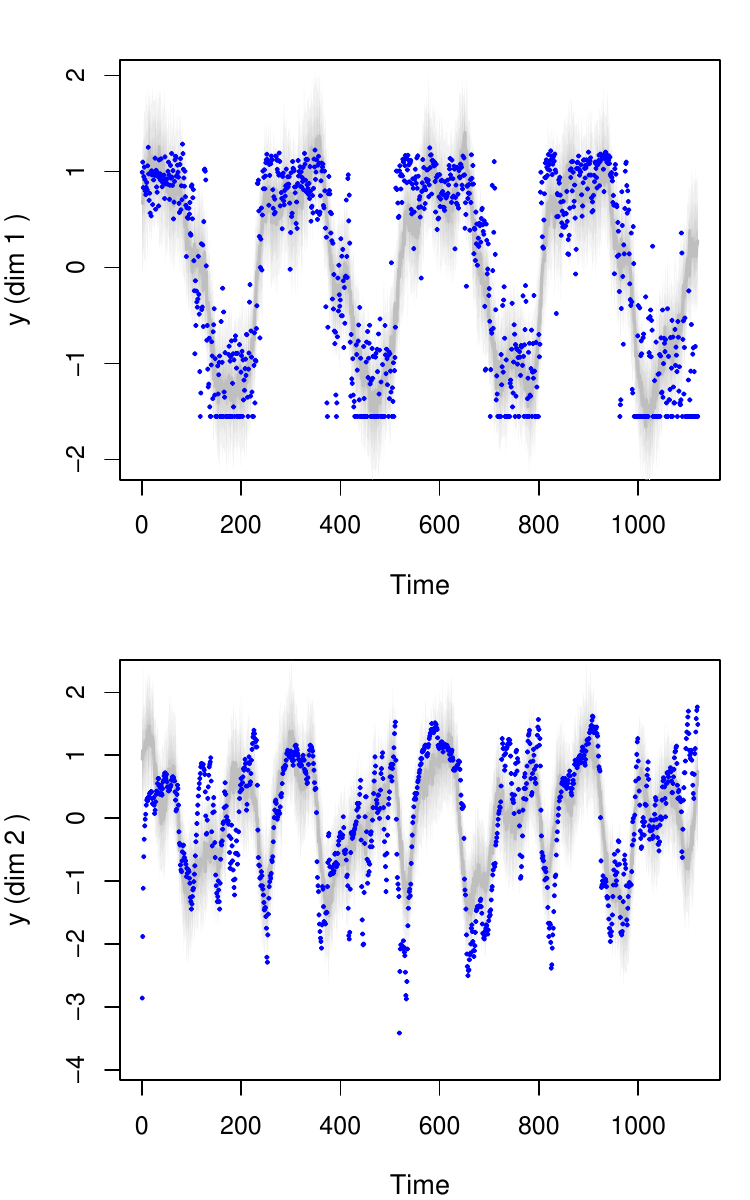}
		\caption{\textbf{Skin Temperature and Activity Monitoring (Bivariate).} Posterior predictive checks for skin temperature (top) and physical activity (bottom). Each panel shows the observed time series along with 30 posterior predictive draws.}
		\label{fig:postpred_subjects_bivar}
	\end{figure}

	To assess the model's ability to reproduce the temporal dynamics of wearable data, we conducted posterior predictive checks for both real-data applications presented in Section~5. Figure~\ref{fig:postpred_subjects_univar} shows results for several individuals with partial-onset seizures, where the model was applied to univariate wrist actigraphy data. Each panel displays the observed time series along with 30 posterior draws from the model. The predictive samples align closely with the observed data, indicating good model fit and capturing both circadian structure and finer-scale fluctuations. Figure~\ref{fig:postpred_subjects_bivar} presents results from the bivariate application, where activity and skin temperature were jointly modeled for a healthy individual. Posterior predictive trajectories again exhibit strong agreement with the data, capturing the shared and distinct features of the two signals.
	
\clearpage
	\bibliographystyle{apalike} 
	\bibliography{Biblio}       

@article{brown:1998,
  title={Multivariate {B}ayesian variable selection and prediction},
  author={Brown, P.J. and Vannucci, M. and Fearn, T.},
  journal={Journal of the Royal Statistical Society, Series B},
  volume={60},
  number={3},
  pages={627--641},
  year={1998},
}

@book{shumway2000time,
  title={Time series analysis and its applications},
  author={Shumway, Robert H and Stoffer, David S and Stoffer, David S},
  volume={3},
  year={2000},
  publisher={Springer}
}

@book{brockwell1991time,
  title={Time series: theory and methods},
  author={Brockwell, Peter J and Davis, Richard A},
  year={1991},
  publisher={Springer science \& business media}
}

@article{savitsky2011variable,
  title={Variable selection for nonparametric {G}aussian process priors: Models and computational strategies},
  author={Savitsky, Terrance and Vannucci, Marina and Sha, Naijun},
  journal={Statistical science: a review journal of the {I}nstitute of {M}athematical {S}tatistics},
  volume={26},
  number={1},
  pages={130},
  year={2011},
  publisher={NIH Public Access}
}

@article{barbieri2004optimal,
  title={Optimal predictive model selection},
  author={Barbieri, Maria Maddalena and Berger, James O},
  journal = "Annals of Statistics",
  volume = {32}, 
  number = {1},
  pages = {870-897},
  year={2004}
}

@article{grant2020ultradian,
  title={Ultradian rhythms in heart rate variability and distal body temperature anticipate onset of the luteinizing hormone surge},
  author={Grant, Azure D and Newman, Mark and Kriegsfeld, Lance J},
  journal={Scientific reports},
  volume={10},
  number={1},
  pages={20378},
  year={2020},
  publisher={Nature Publishing Group UK London}
}

@article{granados2022brain,
  title={Brain waves analysis via a non-parametric Bayesian mixture of autoregressive kernels},
  author={Granados-Garcia, Guilllermo and Fiecas, Mark and Babak, Shahbaba and Fortin, Norbert J and Ombao, Hernando},
  journal={Computational statistics \& data analysis},
  volume={174},
  pages={107409},
  year={2022},
  publisher={Elsevier}
}

@article{wu2024frequency,
  title={Frequency detection and change point estimation for time series of complex oscillation},
  author={Wu, Hau-Tieng and Zhou, Zhou},
  journal={Journal of the American Statistical Association},
  volume={119},
  number={547},
  pages={1945--1956},
  year={2024},
  publisher={Taylor \& Francis}
}

@book{gelman2007data,
  title={Data analysis using regression and multilevel/hierarchical models},
  author={Gelman, Andrew},
  year={2007},
  publisher={Cambridge university press}
}

@article{huang2018hidden,
  title={Hidden Markov models for monitoring circadian rhythmicity in telemetric activity data},
  author={Huang, Qi and Cohen, Dwayne and Komarzynski, Sandra and Li, Xiao-Mei and Innominato, Pasquale and L{\'e}vi, Francis and Finkenst{\"a}dt, B{\"a}rbel},
  journal={Journal of The Royal Society Interface},
  volume={15},
  number={139},
  pages={20170885},
  year={2018},
  publisher={The Royal Society}
}

@article{hadj2020bayesian,
  title={Bayesian model search for nonstationary periodic time series},
  author={Hadj-Amar, Beniamino and Rand, B{\"a}rbel Finkenst{\"a}dt and Fiecas, Mark and L{\'e}vi, Francis and Huckstepp, Robert},
  journal={Journal of the American Statistical Association},
  volume={115},
  number={531},
  pages={1320--1335},
  year={2020},
  publisher={Taylor \& Francis}
}

@book{percival1993spectral,
  title={Spectral analysis for physical applications},
  author={Percival, Donald B and Walden, Andrew T},
  year={1993},
  publisher={cambridge university press}
}

@article{zou2005regularization,
  title={Regularization and variable selection via the elastic net},
  author={Zou, Hui and Hastie, Trevor},
  journal={Journal of the Royal Statistical Society Series B: Statistical Methodology},
  volume={67},
  number={2},
  pages={301--320},
  year={2005},
  publisher={Oxford University Press}
}

@incollection{bretthorst1988excerpts,
  title={Excerpts from {B}ayesian spectrum analysis and parameter estimation},
  author={Bretthorst, G Larry},
  booktitle={Maximum-Entropy and Bayesian Methods in Science and Engineering: Foundations},
  pages={75--145},
  year={1988},
  publisher={Springer}
}

@article{choudhuri2004bayesian,
  title={Bayesian estimation of the spectral density of a time series},
  author={Choudhuri, Nidhan and Ghosal, Subhashis and Roy, Anindya},
  journal={Journal of the American Statistical Association},
  volume={99},
  number={468},
  pages={1050--1059},
  year={2004},
  publisher={Taylor \& Francis}
}

@article{keil2022recommendations,
  title={Recommendations and publication guidelines for studies using frequency domain and time-frequency domain analyses of neural time series},
  author={Keil, Andreas and Bernat, Edward M and Cohen, Michael X and Ding, Mingzhou and Fabiani, Monica and Gratton, Gabriele and Kappenman, Emily S and Maris, Eric and Mathewson, Kyle E and Ward, Richard T and others},
  journal={Psychophysiology},
  volume={59},
  number={5},
  pages={e14052},
  year={2022},
  publisher={Wiley Online Library}
}

@article{tadesse2021handbook,
  title={Handbook of {B}ayesian variable selection},
  author={Tadesse, Mahlet G and Vannucci, Marina},
  year={2021},
  journal = {Chapman and Hall/CRC.},
  publisher={CRC Press}
}

@article{friedrichs2024seizure,
  title={Seizure Cycles under Pharmacotherapy},
  author={Friedrichs-Maeder, Cecilia and Proix, Timoth{\'e}e and Tcheng, Thomas K and Skarpaas, Tara and Rao, Vikram R and Baud, Maxime O},
  journal={Annals of neurology},
  volume={95},
  number={4},
  pages={743--753},
  year={2024},
  publisher={Wiley Online Library}
}

@article{vieluf2024ultradian,
  title={Ultradian rhythms in accelerometric and autonomic data vary based on seizure occurrence in paediatric epilepsy patients},
  author={Vieluf, Solveig and Cantley, Sarah and Krishnan, Vaishnav and Loddenkemper, Tobias},
  journal={Brain Communications},
  volume={6},
  number={2},
  pages={fcae034},
  year={2024},
  publisher={Oxford University Press US}
}

@article{lau2024sleep,
  title={Sleep--wake behavioral characteristics associated with depression symptoms: findings from the Multi-Ethnic Study of Atherosclerosis},
  author={Lau, Stephen CL and Zhang, Gehui and Rueschman, Michael and Li, Xiaoyu and Irwin, Michael R and Krafty, Robert T and McCall, William V and Skidmore, Elizabeth and Patel, Sanjay R and Redline, Susan and others},
  journal={Sleep},
  volume={47},
  number={4},
  pages={zsae045},
  year={2024},
  publisher={Oxford University Press US}
}

@article{abboud2023actigraphic,
  title={Actigraphic correlates of neuropsychiatric symptoms in adults with focal epilepsy},
  author={Abboud, Mark A and Kamen, Jessica L and Bass, John S and Lin, Lu and Gavvala, Jay R and Rao, Sindhu and Smagula, Stephen F and Krishnan, Vaishnav},
  journal={Epilepsia},
  volume={64},
  number={6},
  pages={1640--1652},
  year={2023},
  publisher={Wiley Online Library}
}

@article{lange2021fourier,
  title={From fourier to koopman: Spectral methods for long-term time series prediction},
  author={Lange, Henning and Brunton, Steven L and Kutz, J Nathan},
  journal={Journal of Machine Learning Research},
  volume={22},
  number={41},
  pages={1--38},
  year={2021}
}

@article{tseng2020fourier,
  title={Fourier-transform-based attribution priors improve the interpretability and stability of deep learning models for genomics},
  author={Tseng, Alex and Shrikumar, Avanti and Kundaje, Anshul},
  journal={Advances in Neural Information Processing Systems},
  volume={33},
  pages={1913--1923},
  year={2020}
}

@article{incremona2019spectral,
  title={Spectral characterization of the multi-seasonal component of the Italian electric load: a LASSO-FFT approach},
  author={Incremona, Alessandro and De Nicolao, Giuseppe},
  journal={IEEE Control Systems Letters},
  volume={4},
  number={1},
  pages={187--192},
  year={2019},
  publisher={IEEE}
}

@inproceedings{ha2019performance,
  title={Performance-influence model for highly configurable software with fourier learning and lasso regression},
  author={Ha, Huong and Zhang, Hongyu},
  booktitle={2019 IEEE International Conference on Software Maintenance and Evolution (ICSME)},
  pages={470--480},
  year={2019},
  organization={IEEE}
}

@article{george1993variable,
  title={Variable selection via {G}ibbs sampling},
  author={George, Edward I and McCulloch, Robert E},
  journal={Journal of the American Statistical Association},
  volume={88},
  number={423},
  pages={881--889},
  year={1993},
  publisher={Taylor \& Francis}
}

@article{mitchell1988bayesian,
  title={Bayesian variable selection in linear regression},
  author={Mitchell, Toby J and Beauchamp, John J},
  journal={Journal of the american statistical association},
  volume={83},
  number={404},
  pages={1023--1032},
  year={1988},
  publisher={Taylor \& Francis}
}

@article{adhyapak2024stability,
  title={Stability and Volatility of Human Rest-Activity Rhythms: Insights from Very Long Actograms (VLAs)},
  author={Adhyapak, Nandani and Abboud, Mark A and Rao, Pallavi SK and Kar, Ananya and Mignot, Emmanuel and Delucca, Gianluigi and Smagula, Stephen F and Krishnan, Vaishnav},
  journal={medRxiv},
  year={2024},
  publisher={Cold Spring Harbor Laboratory Preprints}
}

@article{smagula2024sleep,
  title={Sleep-wake behaviors associated with cognitive performance in middle-aged participants of the Hispanic Community Health Study/Study of Latinos},
  author={Smagula, Stephen F and Zhang, Gehui and Krafty, Robert T and Ramos, Alberto and Sotres-Alvarez, Daniela and Rodakowski, Juleen and Gallo, Linda C and Lamar, Melissa and Gujral, Swathi and Fischer, Dorothee and others},
  journal={Sleep Health},
  volume = {10}, 
  number = {4}, 
  pages = {500-507},
  year={2024},
  publisher={Elsevier}
}

@article{heidelberger1981spectral,
  title={A spectral method for confidence interval generation and run length control in simulations},
  author={Heidelberger, Philip and Welch, Peter D},
  journal={Communications of the Association for Computing Machinery},
  volume={24},
  number={4},
  pages={233--245},
  year={1981},
  publisher={ACM New York, NY, USA}
}

@article{hadj2023bayesian,
  title={Bayesian approximations to hidden semi-Markov models for telemetric monitoring of physical activity},
  author={Hadj-Amar, Beniamino and Jewson, Jack and Fiecas, Mark},
  journal={Bayesian Analysis},
  volume={18},
  number={2},
  pages={547--577},
  year={2023},
  publisher={International Society for Bayesian Analysis}
}

@article{komarzynski2018picado,
  title={Picado: a mobile e-health platform for real-time monitoring of circadian dynamics of symptoms and physiological functions in cancer patients},
  author={Komarzynski, Sebastien and Huang, Qi and Innominato, Pasquale F and Bouchahda, Mohamed and Bilski, Karine and Ulusakarya, Ayhan and Beau, Jacqueline and Mocquery, Marion and Adams, Jason and Karaboue, Abdesslam and others},
  journal={NPJ Digital Medicine},
  volume={1},
  number={1},
  pages={1--10},
  year={2018},
  publisher={Nature Publishing Group}
}

@article{tibshirani1996regression,
  title={Regression shrinkage and selection via the lasso},
  author={Tibshirani, Robert},
  journal={Journal of the Royal Statistical Society Series B: Statistical Methodology},
  volume={58},
  number={1},
  pages={267--288},
  year={1996},
  publisher={Oxford University Press}
}

@article{fan2001variable,
  title={Variable selection via nonconcave penalized likelihood and its oracle properties},
  author={Fan, Jianqing and Li, Runze},
  journal={Journal of the American statistical Association},
  volume={96},
  number={456},
  pages={1348--1360},
  year={2001},
  publisher={Taylor \& Francis}
}

@article{zhang2010nearly,
  title={Nearly unbiased variable selection under minimax concave penalty},
  author={Zhang, Cun-Hui},
  journal = {Annals of Statistics},
  volume={38},
  number={2},
  pages={894–-942},
  year={2010}
}

@article{de2024state,
  title={State of the science and recommendations for using wearable technology in sleep and circadian research},
  author={De Zambotti, Massimiliano and Goldstein, Cathy and Cook, Jesse and Menghini, Luca and Altini, Marco and Cheng, Philip and Robillard, Rebecca},
  journal={Sleep},
  volume={47},
  number={4},
  pages={zsad325},
  year={2024},
  publisher={Oxford University Press US}
}

@article{goh2019episodic,
  title={Episodic ultradian events—ultradian rhythms},
  author={Goh, Grace H and Maloney, Shane K and Mark, Peter J and Blache, Dominique},
  journal={Biology},
  volume={8},
  number={1},
  pages={15},
  year={2019},
  publisher={MDPI}
}

@article{grant2018evidence,
  title={Evidence for a coupled oscillator model of endocrine ultradian rhythms},
  author={Grant, Azure D and Wilsterman, Kathryn and Smarr, Benjamin L and Kriegsfeld, Lance J},
  journal={Journal of biological rhythms},
  volume={33},
  number={5},
  pages={475--496},
  year={2018},
  publisher={SAGE Publications Sage CA: Los Angeles, CA}
}

@article{cadonna2019bayesian,
  title={Bayesian spectral modeling for multiple time series},
  author={Cadonna, Annalisa and Kottas, Athanasios and Prado, Raquel},
  journal={Journal of the American Statistical Association},
  year={2019},
  volume = {114}, 
  number = {528},
  pages = {1838-1853},
  publisher={Taylor \& Francis}
}

@article{hu2023fast,
  title={Fast Bayesian inference on spectral analysis of multivariate stationary time series},
  author={Hu, Zhixiong and Prado, Raquel},
  journal={Computational Statistics \& Data Analysis},
  volume={178},
  pages={107596},
  year={2023},
  publisher={Elsevier}
}

\label{lastpage}

\end{document}